\pdfoutput=1
\documentclass[10pt,onecolumn,conference]{IEEEtran}

\usepackage{amsmath,amsfonts,amssymb}
\usepackage{graphicx}
\usepackage{cite}
\usepackage{multirow}
\usepackage{algorithm}
\usepackage{algorithmic}
\usepackage[center]{caption}
\usepackage{amsthm}
\usepackage{amsmath,amsfonts}
\usepackage{amssymb}
\usepackage{enumerate,letltxmacro}
\usepackage{graphicx}
\usepackage{cite}
\usepackage{subfigure}
\usepackage{multirow}
\usepackage{algorithm}
\usepackage{algorithmic}
\usepackage{mathtools}
\usepackage{hyperref}
\usepackage{verbatim}
\usepackage{color, colortbl}

\hypersetup{draft}

\IEEEoverridecommandlockouts

\newtheorem{theorem}{Theorem}
\newtheorem{definition}{Definition}

\newtheorem{corollary}{Corollary}
\newtheorem{proposition}{Proposition}
\newtheorem{example}{Example}
\newtheorem{remark}{Remark}
\newtheorem{case}{Case}

\newcommand{\set}[1]{\mathcal{#1}}

\newcommand{\N}{\set{N}}

\newcommand{\X}{\set{X}}

\title{On Enumerating Distributions for Associated Vectors in the Entropy Space}
\author{\IEEEauthorblockN{Sultan Alam$^{\dagger}$, Satyajit Thakor$^{\dagger}$ and Syed Abbas$^{\ddagger}$}
\IEEEauthorblockN{School of Computing and Electrical Engineering$^{\dagger}$, School of Basic Sciences$^{\ddagger}$\\
Indian Institute of Technology Mandi}
email: \{mohammad\_sultan@students., satyajit@, abbas@\}iitmandi.ac.in
}

\begin{document}

\maketitle
\bibliographystyle{ieeetr}
\begin{abstract}
This paper focuses on the problem of finding a distribution for an associated entropic vector in the entropy space nearest to a given, possibly non-entropic, target vector for random variables with a constraint on alphabet size. We show the feasibility to find distribution for associated vector via a sequence of perturbations in the probability mass function. Then we present an algorithm for numerically solving the problem together with extensions, applications, and comparison with the known results.
\end{abstract}

\section{Introduction}

A fundamental problem in information theory is to determine whether a given vector (point) in the entropy space is entropic or not. This problem is also of practical importance. For instance, a solution to this question can tell us about the existence of a network code with the feasible rate for a given network since if there exists a code with certain rate then there is an entropic point induced by random variables involved and vice versa. Applications to some network information theory  problems are described in \cite{HasSha07a}. If the point does not satisfy an unconstrained information inequality then it is certainly non-entropic. But whether it is entropic can be answered with certainty only if we have the complete characterization of the entropic region (the set of all entropic points). This appears to be an extremely hard problem even for three random variables \cite{ZhaYeu97}. For four or more random variables \cite{ZY98}, even the closure of the entropic region (the set of almost or asymptotically entropic points) is unknown. 

Another approach to address the problem is to determine the existence of a distribution consistent with the given point and if a distribution exists then the point is certainly entropic. In \cite{WalWeb11} an algorithm to determine whether a point is \textit{binary} entropic by checking the feasibility of explicit construction of a consistent binary distribution is given. The algorithm is then used to check binary entropic candidacy of extreme points of an outer bound on normalized \cite[Eq. (5)]{HasSha07a} entropic vectors. The convex hull of entropic extreme points of the outer bound gives an inner bound for normalized entropic vectors. Characterization of similar 
inner bounds has been studied in \cite{Hammer00} via rank functions for vector spaces, in \cite{HasSha07b} via quasi-uniform distributions and in \cite{WalWeb11} via binary distributions.

Also, optimization of functions over the entropic region has been of great interest recently. Well known functions related to Ingleton inequality \cite{Ing71} are Ingleton score \cite{Csi96,DouFreZeg11} and Ingleton violation index \cite{ShaHas10}. While the existence of almost entropic points with lower Ingleton score than the one obtained by the ``four-atom distribution'' is established in \cite{MatCsi16}, the quest for explicit distributions is still on. Ingleton violating entropic points are of special interest since they cannot be induced by vector spaces and subspaces or Abelian groups  and subgroups \cite{Ing71,Cha07}. Moreover, characterization of Ingleton violating almost entropic points would give the \textit{complete} characterization of the closure of entropic region for four random variables.  Ingleton violating entropic points are studied in \cite[Theorem 4]{Hammer00} via an explicit construction, in \cite{MaoThiHas17,BosNan12,Nan15} via construction of groups, in \cite{ShaHas10} via quasi-uniform distributions, in \cite{DouFreZeg11,Liu16} via optimization over distributions with small number of atoms.

A generalization of binary entropic candidacy problem is to determine whether given a vector $\mathbf{h}$, does the ray $E_{\mathbf{h}} \triangleq \{a\mathbf{h}: a >0\}$ contain an entropic point or almost entropic point for a given alphabet size for random variables. This general setup involving a ray is independent of alphabet size and hence from a practical viewpoint, it is more important. Whether a ray contains an entropic point(s) cannot be determined with certainty using the algorithm in \cite{WalWeb11}. An even further generalization of the problem is: \textit{for random variables with a constraint on alphabet size, find a ``nearest'' entropic point to a given ray and associated distribution}. 
In this paper, we focus on addressing this generalized problem.\footnote{
This problem is a generalization for \textit{rays} of ``realizing entropy vectors'' mentioned as a future work in the PhD dissertation \cite{Sha11}.}

In Section \ref{sec:MainResults}, we present a few properties of alphabet constrained entropic sets and define normalized distance. Based on these properties and the notion of normalized distance, in Section \ref{sec: Algorithm}, a randomized local search algorithm to find distributions corresponding to a point nearest to the given target point is given. In Section \ref{sec: Algorithm Extensions and Comparison}, we present extensions of the algorithm, compare the numerical results and discuss a few applications. Conclusion and future work is discussed in Section \ref{sec: Conclusion and Future work}.

\newcommand{\cl}[1]{{\text{cl}({#1})}}

\section{Entropic Sets and Normalized Distance}\label{sec:MainResults}
\subsection{Alphabet Constrained Entropic Sets}
Consider a random vector $X_{\N}=(X_1,\ldots,X_n)$ with the index set $\N=\{1,\ldots,n\}$ over a finite alphabet $\X\triangleq \X_1\times\ldots\times\X_n$ where $\X_i$ is the alphabet of $X_i$. Its probability mass function (pmf) is denoted by the vector $\mathbf{p}=[p(\mathbf{x}), \mathbf{x} \in \X]$. Let $\mathcal P$ be the set of all possible pmfs for the random vector. That is,
\begin{align}
\mathcal P =\left\{\mathbf{p}\in \mathbb{R}^{|\X|}: p(\mathbf{x})\in [0,1],\sum_{\mathbf{x}\in\X}p(\mathbf{x})=1\right\}.
\end{align}

Note that the set $\mathcal P$ is closed, connected and bounded. For a given $\mathbf{p} \in \mathcal P$, let $\mathbf{p}(X_{\alpha}), \emptyset \neq \alpha \subseteq \N$ be the marginal pmfs and let $\mathcal P(X_{\alpha})$ be the set of all possible marginal pmfs for each nonempty $\alpha \subseteq \N$. Also note that $\mathcal P(X_{\alpha})$ is a projection of $\mathcal P$ on to certain coordinates and hence this set too is closed and connected.
We refer a change in quantities as a perturbation. The size of perturbation in pmf is defined as follows.
\begin{definition}[Perturbation size]\label{def:Perturbation size}
The size of perturbation between distributions $\mathbf{p}, \mathbf{p}' \in \mathcal P$ for random vector $X_\N$ is
\begin{align}
\sum_{\mathbf{x}\in \X_1\times\ldots\times\X_n}|p(\mathbf{x})-p'(\mathbf{x})|.
\end{align}
\end{definition}

\begin{remark}
Perturbation size defined above for discrete distributions is the same as the total variation distance (see, e.g., \cite{Ver14}, \cite{HoYeu09} for the definition).
\end{remark}

\begin{corollary}\label{cor:1}
A small perturbation in $\mathbf{p}$ results in a small perturbations in the marginals $\mathbf{p}(X_{\alpha})$. That is,
\begin{align*}
\sum_{\mathbf{x}_{\alpha}}|p(\mathbf{x}_{\alpha})-p'(\mathbf{x}_{\alpha})|
&=\sum_{\mathbf{x}_{\alpha}}\left\vert\sum_{\mathbf{x}_{\alpha}^c}\left(p(\mathbf{x}_{\alpha},\mathbf{x}_{\alpha}^c)-p'(\mathbf{x}, \mathbf{x}_{\alpha}^c)\right)\right\vert\\
&\leq\sum_{\mathbf{x}\in \X_1\times\ldots\times\X_n}|p(\mathbf{x})-p'(\mathbf{x})|.
\end{align*}
\end{corollary}

\begin{definition}[Alphabet constrained entropic set]
Entropic set for a given random vector $X_\N$ over a finite alphabet $\X=\X_1\times\ldots\times\X_n$ is the set of all entropic points and is denoted 
\begin{align}
\Gamma^*_{n,\mathcal X}\triangleq \{\mathbf{h}=[h_{\alpha},\emptyset \neq \alpha\subseteq N]\in \mathbb{R}^{2^n-1}: \mathbf{p}\in \mathcal P\}
\end{align}
where $h_{\alpha} \triangleq h(\mathbf{p}(X_{\alpha}))$ and $h$ is the Shannon entropy function.
\end{definition}

For random vector $(X_1,\ldots,X_n)$, the closure of entropic region (or set) without alphabet constraints is denoted $\overline{\Gamma}^*_n$ \cite{ZhaYeu97}. A well known outer bound on $\overline{\Gamma}^*_n$ is defined by Shannon-type inequalities and is denoted $\Gamma_n$ \cite{ZhaYeu97}. The Shannon-type inequalities are the polymatroidal axioms \cite{Fuj78} and hence any point satisfying the inequalities (and hence in $\Gamma_n$) is also called a polymatroid (the entropy function is a polymatroid function).

The following corollary suggests that a small perturbation in a distribution results in a distribution such that the distance between corresponding entropy points too is small. It may be viewed as a generalization of continuity property of scalar entropy function to entropy vectors.

\begin{corollary}
For $X_\N$, a perturbation in $\mathbf{p}$ results in a new pmf $\mathbf{p}'$. 
As the perturbation size $\delta \rightarrow 0$, the Euclidean distance between $\mathbf{h}$ and $\mathbf{h}'$ also diminishes.
\end{corollary} 
\begin{IEEEproof}
Note that, as $\delta \rightarrow 0$, we have $\mathbf{p}' \rightarrow \mathbf{p}$ and by Corollary \ref{cor:1}, $\mathbf{p}'(X_{\alpha}) \rightarrow \mathbf{p}(X_{\alpha})$ for all non-empty $\alpha \subseteq \N$. Since the entropy function $h$ is continuous for finite alphabet\footnote{The entropy function is discontinuous for countably infinite alphabet size\cite{HoYeu09}.} distributions, ${h}(\mathbf{p}'(X_{\alpha})) \rightarrow{h}(\mathbf{p}(X_{\alpha}))$ 
 and hence $\| \mathbf{h}' -\mathbf{h} \|\rightarrow 0$.
\end{IEEEproof} 
\begin{proposition}\label{prop:1}
$\Gamma^*_{n,\mathcal X}$, for a random vector of size $n$ over finite alphabet $\X$, is a closed connected set.
\end{proposition} 
\begin{IEEEproof}
${\Gamma^*}_{n,\mathcal X}$ is connected since $\mathcal P$ is connected and the continuous image $\mathbf{h}(\mathcal P)$ of this connected set is connected. Let $\mathbf{h}$ be a vector in $\overline{\Gamma}^*_{n,\mathcal X}$. Then there exists a sequence of vectors $\mathbf{h}^{(m)}$ in $\Gamma^*_{n,\mathcal X}$ such that $\mathbf{h}^{(m)}\rightarrow \mathbf{h}$. Hence there exists a sequence of distributions $\mathbf{p}^{(m)}$ such that $\mathbf{h}(\mathbf{p}^{(m)})=[h(\mathbf{p}^{(m)}({X_\alpha})), \emptyset \neq \alpha \subseteq \mathcal N]$ is $\mathbf{h}^{(m)}$. Now, since $\mathcal P$ is closed and bounded implies $\mathbf{p}^{(m)} \rightarrow \mathbf{p}$ where $\mathbf{p}\in \mathcal P$ and continuity of entropy function implies $\mathbf{h}(\mathbf{p}^{(m)}) \rightarrow \mathbf{h}(\mathbf{p})$. Thus, $\mathbf{h}=\mathbf{h}(\mathbf{p})$ and hence is in $\Gamma^*_{n,\mathcal X}$, which implies $\Gamma^*_{n,\mathcal X}$ is a closed set.
\end{IEEEproof}

An implication of Proposition \ref{prop:1} is that, for any given distribution for a random vector over a given alphabet and a given vector in the entropic set, there exists a sequence of perturbations such that the distribution converges to a distribution corresponding to the given entropic vector. 
\subsection{Normalized Distance}

\begin{definition}[Normalized distance]
Consider two polymatroids $\mathbf{x,y} \in \Gamma\setminus \{\mathbf{0}\}$. The normalized distance $d_{\text{\emph{norm}}}(\mathbf{x}, \mathbf{y})$ of $\mathbf{x}$ from $\mathbf{y}$ is the distance between $\mathbf{x}$ and $\mathbf{y}' \triangleq \inf_{\mathbf{y}^*\in E_{\mathbf{y}}}\Vert \mathbf{x}-\mathbf{y}^*\Vert$ divided by the norm of $\mathbf{y}'$. That is,
\begin{align}
 d_{\text{\emph{norm}}}(\mathbf{x}, \mathbf{y})=\frac{\|\mathbf{x}-\mathbf{y}'\|}{\|\mathbf{y}'\|}.
\end{align} 
\end{definition}

Note that, the normalized distance is the tangent of the angle $\theta_{\mathbf{x},\mathbf{y}}$ between the rays defined by the given polymatroids $\mathbf{x},\mathbf{y}$ and hence it is commutative. The distance is ``normalized'' in a sense that it remains the same for any two points on given polymatroidal rays. That is,
$$ \mathbf{x}, \mathbf{x}' \in E_{ \mathbf{x}}, \mathbf{y}, \mathbf{y}' \in E_{ \mathbf{y}} \Rightarrow d_{\text{norm}}(\mathbf{x}, \mathbf{y})= d_{\text{norm}}(\mathbf{x}', \mathbf{y}').$$
It is undefined when $\mathbf{x}$ and $\mathbf{y}$ are orthogonal, $\mathbf{x}\cdot \mathbf{y}=0$, but it is well-defined for polymatroids since no rays defined by two polymatroids are orthogonal (verification is straightforward).

By definition, $d_{\text{norm}}(\mathbf{x}, \mathbf{y})\geq0$. Moreover, we define the equivalence relation as $\mathbf{x} \sim \mathbf{y}$ if $E_\mathbf{x}=E_\mathbf{y}$ and hence equivalence class for a given point $\mathbf{x}$ is $E_\mathbf{x}$. Then, we have 
$$d_{\text{norm}}(\mathbf{x}, \mathbf{y})=0 \Rightarrow E_{ \mathbf{x}}=E_{ \mathbf{y}}, \mathbf{x} \sim \mathbf{y}.$$
But, we note that $d_{\text{norm}}:\Gamma\setminus \{\mathbf{0}\}\times\Gamma\setminus \{\mathbf{0}\}\mapsto\mathbb{R}_+$ is \textit{not} a true metric since it fails to satisfy the triangle inequality
$$d_{\text{norm}}(\mathbf{x}, \mathbf{y})+d_{\text{norm}}(\mathbf{y}, \mathbf{z})\geq d_{\text{norm}}(\mathbf{x}, \mathbf{z})$$
in the range $ \theta_{\mathbf{x},\mathbf{y}}+\theta_{\mathbf{y},\mathbf{z}}\in [0,\pi/2]$ (however, $\tan \theta$ is strictly increasing for $ \theta\in [0,\pi/2]$). Normalized distance has an interesting property: If $d_{\text{norm}}(\mathbf{x}, \mathbf{y'})<d_{\text{norm}}(\mathbf{x}, \mathbf{y})$ then $E_\mathbf{y'}$ is strictly inside the cone
$$C(\mathbf{x},\mathbf{y})\triangleq\{E_\mathbf{y''}: d_{\text{norm}}(\mathbf{x}, \mathbf{y}'')\leq d_{\text{norm}}(\mathbf{x}, \mathbf{y})\}$$
and hence for a sequence of polymatroids $\mathbf{y}^{(m)}$ such that $\mathbf{y}^{(m)}\rightarrow \mathbf{x}$ we have $C(\mathbf{x},\mathbf{y}^{(m)})\rightarrow E_\mathbf{x}$ as $m\rightarrow \infty$.

\begin{theorem}\label{thm:2}
Consider any $\mathbf{p, h}$ and a given vector $\mathbf{h}_t$. If there exists no perturbation $\mathbf{h}'$ of $\mathbf{h}$ such that it can result in $d_{\text{\emph{norm}}}(\mathbf{h'}, \mathbf{h}_t)< d_{\text{\emph{norm}}}(\mathbf{h}, \mathbf{h}_t)$ then either $\mathbf{h} \in E_{\mathbf{h}_t}$ or $\mathbf{h}$ is at the boundary of the entropic set (an exterior).
\end{theorem} 
\begin{IEEEproof}
Note that if $d_{\text{norm}}(\mathbf{h}, \mathbf{h}_t)=0$ then $\mathbf{h} \in E_{\mathbf{h}_t}$. 
Now assume that $\mathbf{h}$ is in interior of the entropic set and $d_{\text{norm}}(\mathbf{h}, \mathbf{h}_t)>0$ then it is sufficient to show that there always exists a perturbation $\mathbf{h}'$ of $\mathbf{h}$ such that it can result in $d_{\text{norm}}(\mathbf{h'}, \mathbf{h}_t)< d_{\text{norm}}(\mathbf{h}, \mathbf{h}_t)$. Now, for any $\epsilon>0$ there exists entropic $\mathbf{h'}$ such that  $d_{\text{norm}}(\mathbf{h'}, \mathbf{h}_t)< d_{\text{norm}}(\mathbf{h}, \mathbf{h}_t)+\epsilon$ since the entropic set is closed and connected. Hence, by letting $\epsilon \rightarrow 0$, we have $d_{\text{norm}}(\mathbf{h'}, \mathbf{h}_t)\leq d_{\text{norm}}(\mathbf{h}, \mathbf{h}_t)$ where equality holds only when $E_{\mathbf{h}}=E_{\mathbf{h}'}$. When $E_{\mathbf{h}}\neq E_{\mathbf{h}'}$, there always exists $\mathbf{h}'$ in the $\epsilon$-ball of $\mathbf{h}$ for which $d_{\text{norm}}(\mathbf{h'}, \mathbf{h}_t)< d_{\text{norm}}(\mathbf{h}, \mathbf{h}_t)$.
\end{IEEEproof}

\begin{proposition}\label{prop:2}
Given a complete characterization of $\Gamma^*_{n,\mathcal X}$, a distribution for a nearest (by $d_{\text{\emph{norm}}}$) entropic vector  to a given target vector $\mathbf{h}_t$ can be obtained with arbitrary precision by starting at any distribution and iteratively perturbing the distribution with perturbations in the range $(0,\epsilon]$ for any $\epsilon>0$.
\end{proposition} 
\begin{IEEEproof}
If the complete characterization of the entropic set $\Gamma^*_{n,\mathcal X}$ is known, it is feasible to find $\mathbf{h}'=\inf_{\mathbf{h}\in \Gamma^*_{n,\mathcal X}}d_{\text{norm}}(\mathbf{h},\mathbf{h}_t)$. Now, since the entropic set is connected, Proposition \ref{prop:1}, one can start from any distribution and choosing appropriate intermediate target points (to avoid perturbations leading to some another boundary point of the entropic set, see Theorem \ref{thm:2}), can obtain a distribution consistent with $\mathbf{h}'$ by trying random perturbations in the range $(0,\epsilon]$ for any $\epsilon>0$.
\end{IEEEproof} 

On the other hand, if we have some distributions then conic hull of respective entropic points gives an inner bound for $\overline{\Gamma}^*_n$. Also, in Proposition \ref{prop:2} it is sufficient to choose (to ensure non-covergence at a boundary point other than $\mathbf{h}'$) intermediate target points such that for consecutive target points $\mathbf{h}_i$, $\mathbf{h}_{i+1}$, $C(\mathbf{h}_i$, $\mathbf{h}_{i+1})\cap C(\mathbf{h}_{i+1}$, $\mathbf{h}_{i})\subset \Gamma^*_{n,\mathcal X}$ if such points exist, though it is not necessary. 

\begin{remark}
In practice, it may be of interest to find whether a given vector is entropic. This may be referred to as ``realizing entropy vectors'' mentioned in \cite{Sha11}. For a give alphabet size, this problem can be answered using the procedure in \cite{WalWeb11}. Apart from the procedure in \cite{WalWeb11}, the procedure in \cite{Kou0?0} for entropy scalars may be extended for realizing entropy vectors.  
In this paper, we focus on whether there exists an entropic point on a ray defined by a given vector. The procedure \cite{WalWeb11} cannot guarantee an answer for this problem. The motivation to focus on rays rather than vectors is 
that at the boundary of the almost entropic region there may exist rays with non-entropic vectors (see \cite[Chapter 15]{Yeu08} for details). Hence, the existence of a non-entropic vector dose not necessarily mean that the ray defined is not in  the almost entropic region. 
\end{remark}

\subsection{Additional Information Equality Constraints}

Now we discuss entropic sets with constraints on distributions in addition to alphabet size, such sets are of practical importance since a communication system usually induces constraints on the distributions of random variables involved. Moreover, in many cases such constraints on distributions translate into linear entropic equalities (e.g., functional dependence, conditional or unconditional independence). For instance, network coding capacity region \cite{YanYeuZha12} for alphabet constrained system will simply be the projection of the intersection of the entropic set with network induced linear information constraints onto the coordinates associated with source entropies.

\begin{corollary}
The entropic set $\Gamma^*_{n,\X}$ constrained by linear entropic constraints can, in general, be a disconnected set.
\end{corollary}
\begin{IEEEproof}
The corollary follows from the example of the unique extreme ray of $\Gamma_3$ which contains disconnected entropic points \cite{ZhaYeu97} (also see \cite[Chapter 15]{Yeu08}). Alternatively, since $\Gamma^*_{n,\X}$ is non-convex in general \cite{HasSha07a}, its intersection with linear or non-linear  constraints can result in a disconnected set. 
\end{IEEEproof} 

Note that, Proposition \ref{prop:2} is \textit{not valid} for constrained entropic sets because due to disconnectedness, convergence cannot be guaranteed by starting from any distribution for a point in the constrained entropic set and  employing  perturbations in the range $(0,\epsilon]$ for any $\epsilon>0$. Thus, searching over constrained (and hence in general disconnected) entropic set via arbitrary perturbations may not yield fruitful results. However, given an entropic vector associated with achivable rate vector for a communication system (and hence in a constrained entropic set), we can still conduct a search over $\Gamma^*_{n,\X}$ using perturbations in the range $(0,\epsilon]$ for a distribution consistent with the entropic vector.

\section{Algorithm}\label{sec: Algorithm}
Based on the content in the previous section, we present a randomized local search algorithm, Algorithm 1. For a given target polymatroid $\mathbf{h}_t \in \Gamma_n \setminus \{\mathbf{0}\}$, the algorithm iteratively tries perturbations to find a distribution with strictly less $d_{\text{norm}}$ of associated entropic point from the target point, starting from a distribution $\mathbf{p}_s \in\mathcal{P}(\X)$ on a given alphabet $\X$. Here, we are employing the simplest type of perturbation in which a distribution and its perturbed distribution differ only in two probability mass points.

\begin{algorithm}
\caption{{Find $\mathbf{h}\in \Gamma^*_{n,\X}$ near to $\mathbf{h}_t$ and associated $\mathbf{p}$.}}\label{alg:d-norm}
  \begin{algorithmic}[1]
    \REQUIRE $\mathbf{h}_t, \mathbf{p}_s,\delta, L, M,\epsilon\in(0,1]$
    \ENSURE{$\mathbf{p}_{k}, \mathbf{h}_{k}$}
\STATE$\mathbf{p}_k\leftarrow\mathbf{p}_s$
    \STATE$\mathbf{h}_k \leftarrow h( \mathbf{p}_k)$
    \FOR{$l=1$ to $L$}
    \STATE$\mathbf{p}'_k\leftarrow\mathbf{p}_k$
    \FOR{$m=1$ to $M$}
    \STATE{Choose indexes $i,j$ of $\mathbf{p}_k$ uniformly randomly}
    \STATE{Choose $\lambda \in[0,\epsilon]$ uniformly randomly}
    \STATE{$\mathbf{p}'_k(i) \leftarrow \lambda(\mathbf{p}_k(i)+\mathbf{p}_k(j))$}
    \STATE{$\mathbf{p}'_k(j) \leftarrow (1-\lambda)(\mathbf{p}_k(i)+\mathbf{p}_k(j))$}
    \STATE{$\mathbf{h}'_k\leftarrow h(\mathbf{p}'_k)$}
    \IF{$d_{\text{norm}}(\mathbf{h}'_k,\mathbf{h}_t)< d_{\text{norm}}(\mathbf{h}_k,\mathbf{h}_t)$}
    \STATE{$\mathbf{p}_{k}\leftarrow\mathbf{p}'_k,  \mathbf{h}_{k}\leftarrow h(\mathbf{p}_{k})$, exit for loop}
    \ENDIF
    \ENDFOR
    \IF{$m=M$}
    \RETURN{$\mathbf{p}_{k}, \mathbf{h}_{k}$ and terminate}
    \ELSIF{$d_{\text{norm}}(\mathbf{h}_k,\mathbf{h}_t)\leq \delta$}
    \RETURN{$\mathbf{p}_{k}, \mathbf{h}_{k}$ and terminate}
    \ENDIF
    \ENDFOR
\RETURN{$\mathbf{p}_{k}, \mathbf{h}_{k}$}
  \end{algorithmic}
\end{algorithm}

\begin{proposition}\label{prop:3}
For sufficiently large $L$ and $M$, Algorithm 1 terminates with an output distribution and its entropic vector at least $\delta$-near (in $d_{\text{\emph{norm}}}$) to either the ray defined by the target vector or a boundary point of $\Gamma^*_{n,\X}$ for any given $\delta\geq 0$.
\end{proposition} 
\begin{IEEEproof}
The statement follows by Theorem \ref{thm:2}.
\end{IEEEproof} 

The algorithm can be used to find entropic points (and associated distributions) near to the boundary of $\Gamma^*_{n,\X}$ in various directions by choosing appropriate starting points and the target points. Moreover, we can replace $d_{\text{norm}}$ with Euclidean distance, or another metric, to obtain numerical results for the respective minimization problem.

For the algorithm, letting the initial distribution $\mathbf{p}_s$ as uniform independent distribution is an obvious and easy choice. But, since the entropic vector corresponding to a random vector with independent random variables is at the boundary or a supporting hyperplane of $\overline{\Gamma}^*_{n}$ (and hence, also at the boundary of $\Gamma^*_{n,\X}$), it is not the best choice for every given target vector.  In general, we take the starting distribution as the cetroid of $\Gamma_n$ due to lack of characterization of entropic or almost entropic vectors for $n\geq4$. 
\begin{definition}[Centroid and centroid ray]
A \textit{centroid} $\mathbf{c}$ of a set in an Euclidean space is the mean of the points in the set, similarly \textit{centroid ray} $E_{\mathbf{c}}$ for a pointed cone is the mean of its rays.
\end{definition}

For instance, the centroid ray of $\Gamma_n$ is the mean of its extreme rays and is the ``inner most'' ray of $\Gamma_n$. A more effective choice for initial distribution can be made from the knowledge available about the shape of $\Gamma^*_{n,\X}$ and the orientation of $\mathbf{h}_t$. 

\begin{figure}[htbp]
\centering
  \includegraphics[scale=0.27]{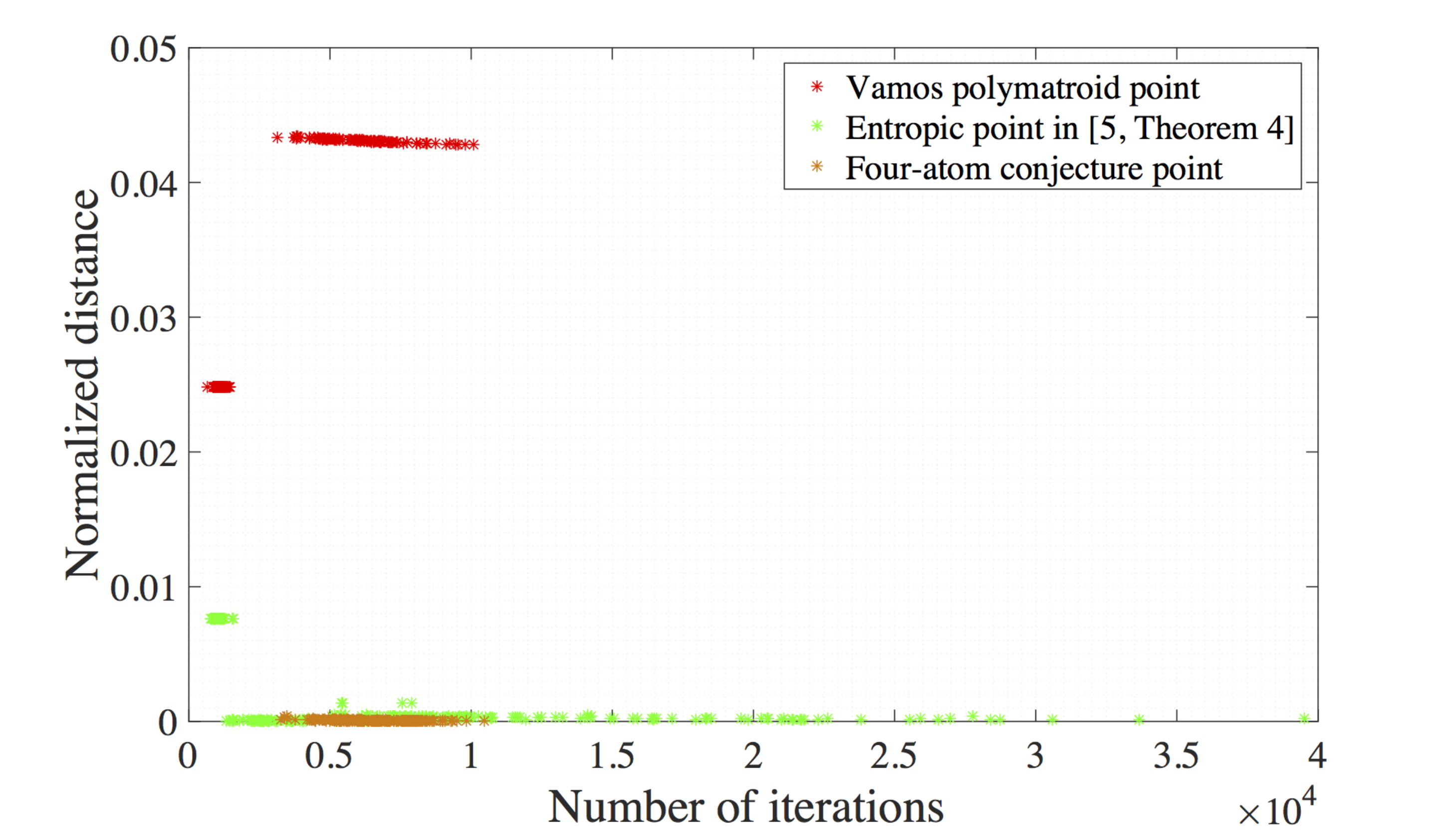}
  \caption{Numerical results for three different target points.}\label{fig:plot}
\end{figure}
Figure \ref{fig:plot} shows the simulation results of Algorithm 1 in terms of normalized distance 
and the number of iterations. An asterisk represents the normalized distance of a vector return by the algorithm. The target points are (1) the V\'{a}mos polymatroid (not an almost entropic point), (2) the entropic point constructed in \cite[Theorem 4]{Hammer00} and (3) the four-atom conjecture point \cite{DouFreZeg11}. The initial point is a point on the centroid ray of the base points of the Ingleton pyramid \cite{DouFreZeg11} (also see the next section) and alphabet size is 2 for each random variable. We note that for entropic vectors, four-atom conjecture point (brown asterisks) and the point in \cite[Theorem 4]{Hammer00} (green asterisks), in most instances the algorithm terminates with a point very close to the target point in $d_{\text{norm}}$. But, in a few instances, the returned points are not very near to the target entropic points. This is due to no-convexity of the entropic set and a particular trajectory (or path or direction) defined by random perturbations. However, as suggested by Theorem \ref{thm:2} and Proposition \ref{prop:3}, the returned points are most likely very near to the boundary of the entropic set. For V\'{a}mos polymatroid, there are two clusters of the return points (red asterisks). 
It is important to note that the return results are dependent on the initial point and the trajectory (which may be specified by defining intermediate target points) and even more consistent results may be obtained by suitable choice of these parameters.
The exact distributions and additional numerical results are made available in the appendix.

\section{Extensions, Comparison and Applications}\label{sec: Algorithm Extensions and Comparison}
We first present a few extensions to optimize different functions over an entropic set. We restrict our attention to examples and numerical results for the random vectors of size four.

We briefly describe the known results for comparison. For $\mathbf{h}$ the Ingleton expression is 
$h_{12}+h_{13}+h_{23}+h_{14}+h_{24}-h_1-h_2-h_{123}-h_{124}-h_{34} \triangleq \Delta_{34}$
and the Ingleton inequality is $\Delta_{34}\geq0$ \cite{Ing71}. There are six distinct Ingleton inequalities via permutation. The Ingleton score for $\mathbf{h}$ is
$\mathbb{I}(\mathbf{h})\triangleq \Delta_{34}/h_{1234}$ \cite{DouFreZeg11,MatCsi16} and the Ingleton violation index is
$\iota(\mathbf{h})\triangleq-\Delta_{34}/\parallel \mathbf{h}\parallel$ \cite{ShaHas10}.
It has been shown in \cite{MatCsi16} that a polymatroid $\mathbf{h}$ can be decomposed into modular $\mathbf{h}^{\text{mod}}$ and tight $\mathbf{h}^{\text{ti}}$ components such that $\mathbf{h}= \mathbf{h}^{\text{mod}} +\mathbf{h}^{\text{ti}}$. Since, modular polymatroids are trivially entropic, the study of entropic region reduces to characterization of tight entropic polymatroids. Also, it has been shown that for entropic polymatroid, its tight component is almost entropic. These new results are used to refute the four-atom conjecture. Moreover, it has been shown that a particular linear transformation, we denote $t(\cdot)$, on an almost entropic tight point yields another almost entropic tight point with better Ingleton score (see \cite{MatCsi16} for details). 

As suggested in the previous section, the algorithm can be extended for a variety of optimization functions. We also extend our perturbation based technique to optimize the functions $\mathbb{I}$ and $\iota$. 
As summarized in Table \ref{tab:1}, we get better and in some cases nearly as good numerical results. So far, we have obtained simulations for only upto alphabet size of 5 (or less  in some cases). Various initial points and a new search approach is used. For instance, to \textit{restrict} the search area, we propose incorporating \textit{alternative hyperplane score reduction} in addition to normalized distance reduction (see the appendix for details about the approach and detailed numerical results).
\begin{table}[!htbp]
\centering
\caption{Comparison of numerical results}
\label{tab:1}
\begin{tabular}{|c|l|l|l|}
\hline
\multirow{2}{*}{Func. Opt.} & \multicolumn{2}{c|}{Known results}                          & \multicolumn{1}{c|}{Our results} \\ \cline{2-4} 
                       & \multicolumn{1}{c|}{Result} & \multicolumn{1}{c|}{Approach} & \multicolumn{1}{c|}{Result ($|\X|$)}      \\ \hline
$\min \mathbb{I}(\mathbf{h})$&  -0.089373 &     Newton's method \cite{DouFreZeg11}  &    -0.089373    ($2^{4}$)     \\ \hline
\multirow{2}{*}{$\max \iota(\mathbf{h})$}     &     0.02761      &   MCMC-Quasi U. \cite{ShaHas10}  & \multirow{2}{*}{0.0281316 ($2^4$)}                \\ \cline{2-3}
                       &    0.02812    &    Empirical Dist. \cite{Sha11}        &                                  \\ \hline
\multirow{2}{*}{$\min \mathbb{I}(t(\mathbf{h}^{\text{ti}}))$} &   -0.09243      &    Various \cite{MatCsi16}  & \multirow{2}{*}{-0.092499 ($5^4$)}               \\ \cline{2-3}
                   &        -0.092500       &    Groups \cite{Nan15}     &                                      \\ \hline                   
\end{tabular}
\end{table}

Now, we present a simple approach to obtain an inner bound for a given polyhedral outer bound on $\overline{\Gamma}^*_n$ with the \textit{same ``description complexity''} but with significant improvement over known inner bounds. Here, by equal description complexity, we mean that the inner bound is expressed using the same number of inequalities as the outer bound. This is extremely important in practice where the interest is to use inner bounds for computational purposes, e.g., computing an inner bound for network coding capacity, and it is desirable that the best possible inner bound is expressed by only certain number of linear inequalities.\footnote{For instance, more than a million points in the pyramid are obtained in \cite{MatCsi16}, and a thousand points in \cite{Liu16} but the use of it for computational purposes is mostly unfeasible due to the high number of linear inequalities in the description.} The simple approach is, for the extreme rays of a given polyhedral outer bound, to obtain nearest entropic rays. The conic hull of these entropic rays delivers an inner bound. To demonstrate the idea, consider the part of $\Gamma_4$ cut-off by reverse Ingleton inequality, such a cone is referred as a pyramid in \cite{DouFreZeg11}. Characterization of $\overline{\Gamma}^*_4$ reduces to the entropic points in this pyramid defined by 15 extreme rays \cite{DouFreZeg11} (actually, the characterization is further reduced to the 11-dimensional sub-cone of tight points \cite{MatCsi16}) and hence we focus on the entropic region in the pyramid. The pyramid is further cut-off by two ZY98 \cite{ZY98} inequalities. This removes one extreme ray (associated with V\'{a}mos polymatroid) and introduces 9 additional extreme rays. This outer bound with 23 extreme rays (or equivalently, 17 hyperplanes) has a relative volume of 98.4568\% \cite{DouFreZeg11}. Our inner bound with the same description complexity has relative volume 43.8026\%. Also, we obtained best inner bound with relative volume 56.2590\% and with least number of extreme rays 285
using a new \textit{grid approach} (see the appendix for details). We emphasis that for numerical results we have only used binary distributions and by allowing higher alphabets the volume of inner bound can be further increased. Inner bounds obtained this way may be further used, for instance, to find achievable rate vectors for network coding.

Also, we find binary entropic points near to 35 extreme rays of the Ingleton cone. The total of 26 points thus obtained are very close to the respective extreme rays and hence approximately binary entropic as it can be confirmed by results in \cite{WalWeb11}. For the remaining 9 extreme rays near binary entropic points are also found  and the numerical results confirm that these 9 extreme rays are not binary entropic as listed in \cite[Table 1]{WalWeb11} (see the appendix for details).

\section{Conclusion and Future Work}\label{sec: Conclusion and Future work}
We presented the notion of $d_{\text{norm}}$ and perturbation based randomized local search algorithm which, in principle, has the potential to deliver numerical inner approximations for alphabet constrained entropic sets. Various extensions of the algorithm are presented and the numerical results are compared to the known results. The application to obtain inner bounds from outer bounds with the same description complexity also signifies the importance of the main results. We note that though the algorithm employs randomized local search, trajectory (or path or direction) can be specified by defining intermediate target points - this feature may not be readily available for many function minimization problems. 

As a future direction, systematic approaches to obtain inner bounds for the entropic region in the pyramid shall be investigated. 
Also, the problem of V\'{a}mos network \cite{DFZ07} capacity will be investigated based on the results in this paper.
\section*{Acknowledgment}
This work is supported by SERB, DST, Government of India, under Extra Mural Scheme SB/S3/EECE/265/2016. We thank the reviewers for the comments and suggestions.
\bibliographystyle{ieeetr}
\bibliography{network}

\appendix
The appendix contains additional numerical results, methods, and approaches.  Note that, the extreme ray can be described by any one point in the extreme ray and hence no distinction is made. Also, ``distance'' and ``normalized distance'' are not distinguished and shall be clear from the context. Most numerical results here are approximate in a sense that a certain number of digits after the decimal point are presented.
The content of this document is organized as follows:
\begin{itemize}
\item Section \ref{sec:1} presents the simulation results for three target points are presented: (1) the V\'{a}mos polymatroid, (2) the entropic point constructed in \cite[Theorem 4]{Hammer00} and (3) the four-atom conjecture point.
\item Section \ref{sec:2a} presents simulation results for optimizing the Ingleton score and Ingleton index.
\item Section \ref{sec:2b} presents a new hyperplane score based approach for normalized distance reduction.
\item Section \ref{sec:2c} presents a systematic procedure to obtain inner bound from known outer bounds on the entropy region.
\item Section \ref{sec:2d} presents a grid based approach to obtain inner bounds. Using this approach we obtain the best known inner bound (with least number of extreme rays) for the entropy region in the pyramid with only 285 extreme rays and with relative volume 56.2590\%.
\item Section \ref{sec:2e} presents simulation results for Ingleton cone extreme rays as target points. We verify that for all binary entropic Ingleton cone extreme rays, the algorithm returns points very near to the extreme rays considering binary distributions. For all other extreme rays of the Ingleton cone, we obtain near binary entropic points. 
\item Section \ref{sec:app1} presents the joint distributions of four random variables with alphabet size 5 each for which the Ingleton score of the associated tight vector is $-.092499$.
\end{itemize}
\subsection{Results for Three different Target Points}\label{sec:1}
Figure \ref{fig:plot} shows the simulation results of Algorithm 1 in terms of normalized distance 
and number of iterations $l$. The target points are (1) the V\'{a}mos polymatroid (not an almost entropic point), (2) the entropic point constructed in \cite[Theorem 4]{Hammer00} and (3) the four-atom conjecture point \cite{DouFreZeg11}. The algorithm is run $200$ times for each target point. The initial point is approximately in (or near to a point in) $E_{\mathbf{c}}$ and alphabet size is 2 for each random variable. We note that for entropic vectors, four-atom conjecture point (brown asterisks) and the point in \cite[Theorem 4]{Hammer00} (green asterisks), in most instances the algorithm terminates with a point very close (in $d_{\text{norm}}$) to the target point. But, in a few instances, the returned points are not very near to the target entropic points. This is due to no-convexity of the entropic set and a particular trajectory (or path or direction) defined by random perturbations. However, as suggested by Theorem 1 and Proposition 3, the returned points are most likely very near to the boundary of the entropic set. For V\'{a}mos polymatroid, there are two clusters of the return points (red asterisks). 
It is important to note that the return results are dependent on the initial point and the trajectory (which may be specified by defining intermediate target points) and even more consistent results may be obtained by suitable choice of these parameters.

(1) Letting V\'am{o}s polymatroid as the target point and a point on the centroid ray of the base points of the pyramid as the starting point, we obtained $200$ points returned by the algorithm. Of them, the nearest point is  
$$[0.9258 \;0.9257 \;1.5842 \;0.9998 \;1.4807 \;1.4825 \;1.9531 \;0.9998 \;1.4825 \;1.4807 \;1.9531 \;1.9084 \;1.9607 \;1.9607 \;1.9829]$$
with Ingleton score $= -0.078348$, and the normalized distance  $= 0.024821$ from the target, and is attained by the following joint distribution of four random variables with alphabet $\{0,1\}$ each:
\begin{equation*}
\begin{split}
&p_{0000}=0.001644220317494, p_{1100}=0.336225764806443, p_{0010}= 0.003087140924877, p_{1010}=0.163408647228728,\\ &p_{1110}=0.003089948192363, p_{0001}=0.000000004117487, p_{1001} = 0.154625194533945, p_{1101}= 0.000000000000719, \\&p_{0011}= 0.336271989652690, p_{1111} =0.001647090225254,  
\end{split}
\end{equation*} 
and else zero. 

(2) Letting Construction in \cite[Theorem 4]{Hammer00} as the target point and a point on the centroid ray of the base points of the pyramid as the starting point, we obtained $200$ points returned by the algorithm. Of them, the nearest point is 
$$[1.0000 \;1.0000 \;1.9544 \;1.0000 \;1.8113 \;1.8113 \;2.6225 \;1.0000 \;1.8113 \;1.8113 \;2.6225 \;2.0000 \;2.5000 \;2.5000 \;2.9056]$$
with Ingleton score $= -0.015631$, and normalized distance  $= 9.1499\cdot 10^{-06}$ from the target, and is attained by the following joint distribution of four random variables with alphabet $\{0,1\}$ each:
\begin{equation*}
\begin{split}
&p_{0000}=0.000000000000089, p_{0100}= 0.000004110164949, p_{1000}= 0.000000000057158, p_{1100}=0.247468049947254,\\ &p_{0010}= 0.029630844445635, p_{0110} =0.093438275305339,  p_{1010}=0.094132895951530, p_{1110}=0.032959081933003,\\ &p_{0001}=0.029584458949872, p_{0101}=0.093357834010918, p_{1001} = 0.094049290457961, p_{1101}= 0.032908063729594,\\ &p_{0011}= 0.252466526959316, p_{0111} =0.000000568087382,
\end{split}
\end{equation*} 
and else zero. 

(3) Letting four-atom conjecture point as the target point and a point on the centroid ray of the base points of the pyramid as the starting point, we obtained $200$ points returned by the algorithm. Of them, the nearest point is  
$$[ 0.9345 \;0.9345 \;1.5811 \;1.0000 \;1.4401 \;1.4401 \;1.8802 \;1.0000 \;1.4401 \;1.4401 \;1.8802 \;1.8801 \;1.8802 \;1.8802 \;1.8803]$$
with Ingleton score $= -0.089355$, and normalized distance  $= 2.1080\cdot 10^{-05}$ from the target, and is attained by the following joint distribution of four random variables with alphabet $\{0,1\}$ each:
\begin{equation*}
\begin{split}
&p_{1100}=0.350477529019748, p_{0110}= 0.149490839096973, p_{1010}= 0.000000599702761, p_{0001}=0.000004708298428,\\ &p_{0101}= 0.149527616553994, p_{1001} =0.000000000131247,  p_{1101}=0.000004898377650, p_{0011}=0.350493808818975,\\ &p_{1011}=0.000000000000224,
\end{split}
\end{equation*} 
and else zero. 

\begin{figure}[htbp]
\centering
  \includegraphics[scale=0.27]{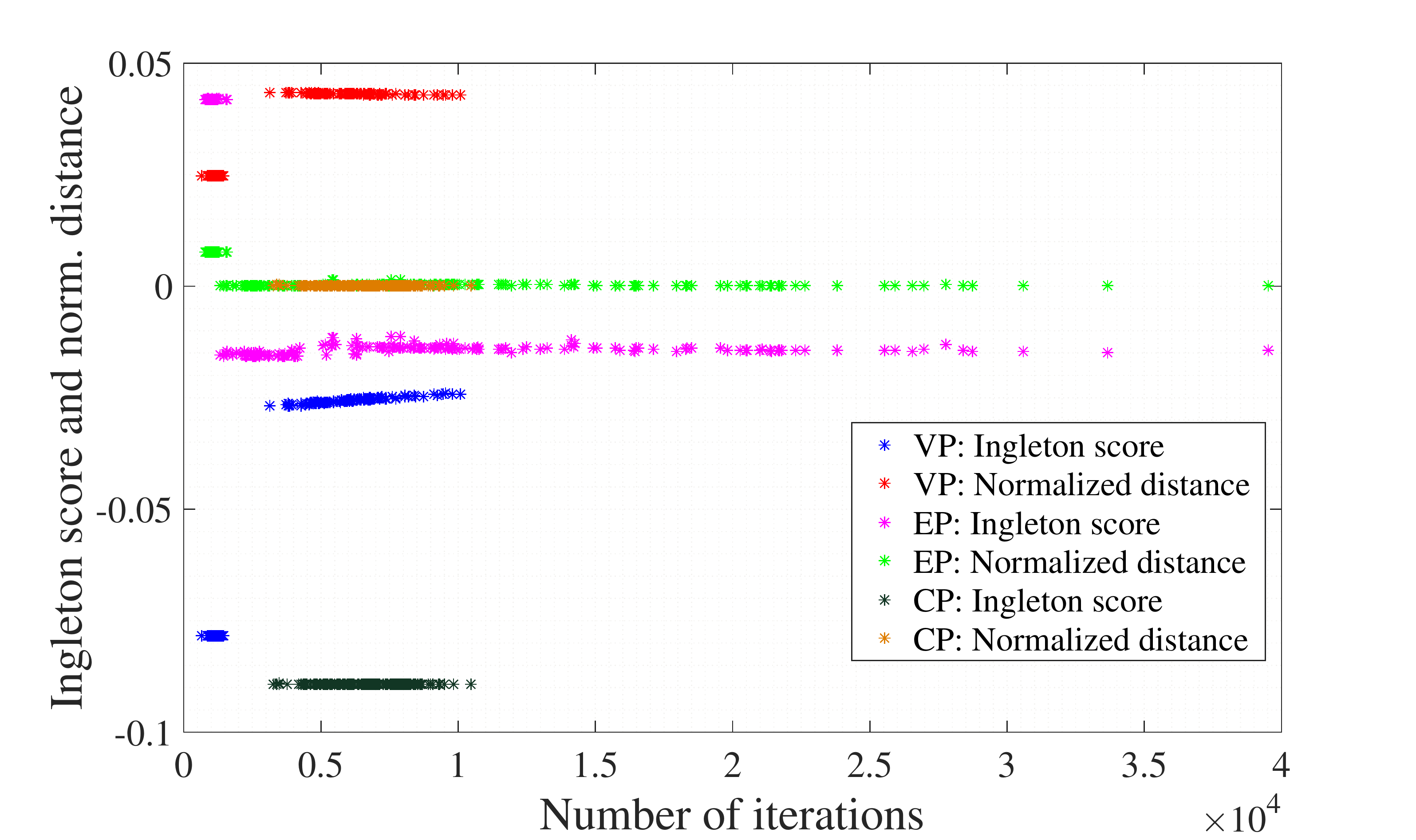}
  \caption{Numerical results with Ingleton scores. Vamos polymatroid point (VP), Entropic point in [5, Theorem 4] (EP), Four-atom conjecture point (CP).}\label{fig:plot1}
\end{figure}

Extended plot with Ingeton scores for all three points is given in Figure \ref{fig:plot1}.

\subsection{Extensions, Comparison and Applications}
\subsubsection{Algorithm Extensions and Comparison}\label{sec:2a}
We have extended  our perturbation based technique to optimize the functions $\mathbb{I}$ and $\iota$. 
As summarized in Table \ref{tab:1}, we get better and in some cases nearly as good numerical results. We have obtained simulations for only upto alphabet size of 5 (or less  in some cases). Below are the details of the numerical results.

\textbf{Minimization of $\mathbb{I}(\mathbf{h})$}: In the algorithm, initial distribution is for a point approximately in centroid ray of $\Gamma_4$ for binary random variables. We have minimized the Ingleton score. The minimum Ingleton score $\mathbb{I}(\mathbf{h}) = -0.089373$, is attained by the following joint distribution:
\begin{equation*}
\begin{split}
&p_{1100} = 0.350527620352674, p_{0010}= 0.000000011646893, p_{1010}=   0.149583628030039, p_{1110}=0.000000000000452, \\&p_{0001}=0.000000000000004, p_{1001}=0.149531459224755, p_{0011}=0.350357280745183,
\end{split}
\end{equation*}
and else zero.

\begin{figure}[htp!]
 	\centering
 	\subfigure[\label{IngAndViolationIndexVsIteration}]{\includegraphics[scale=.27]{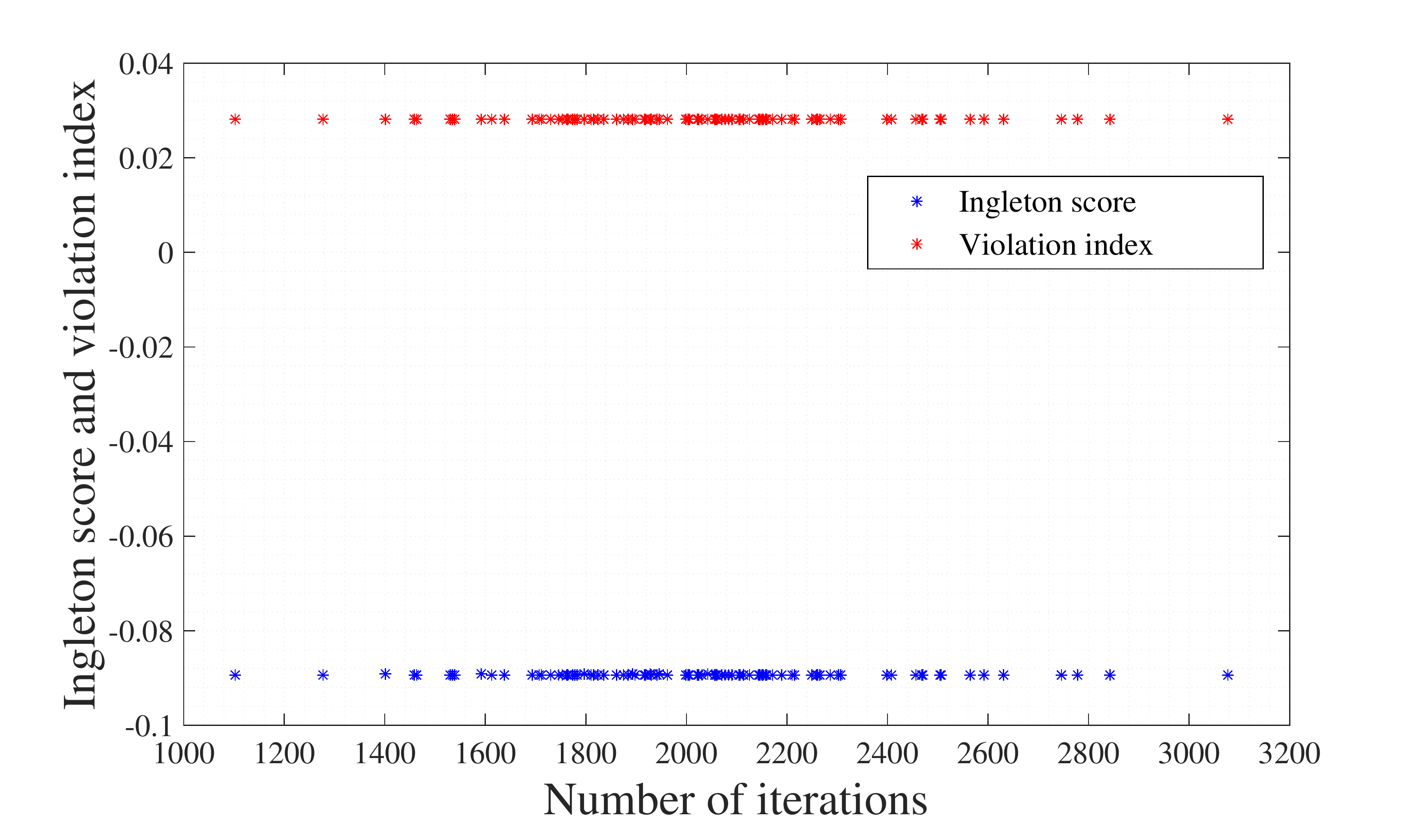}}
 	\subfigure[\label{OurViolationIndexFor2As}]{\includegraphics[scale=.27]{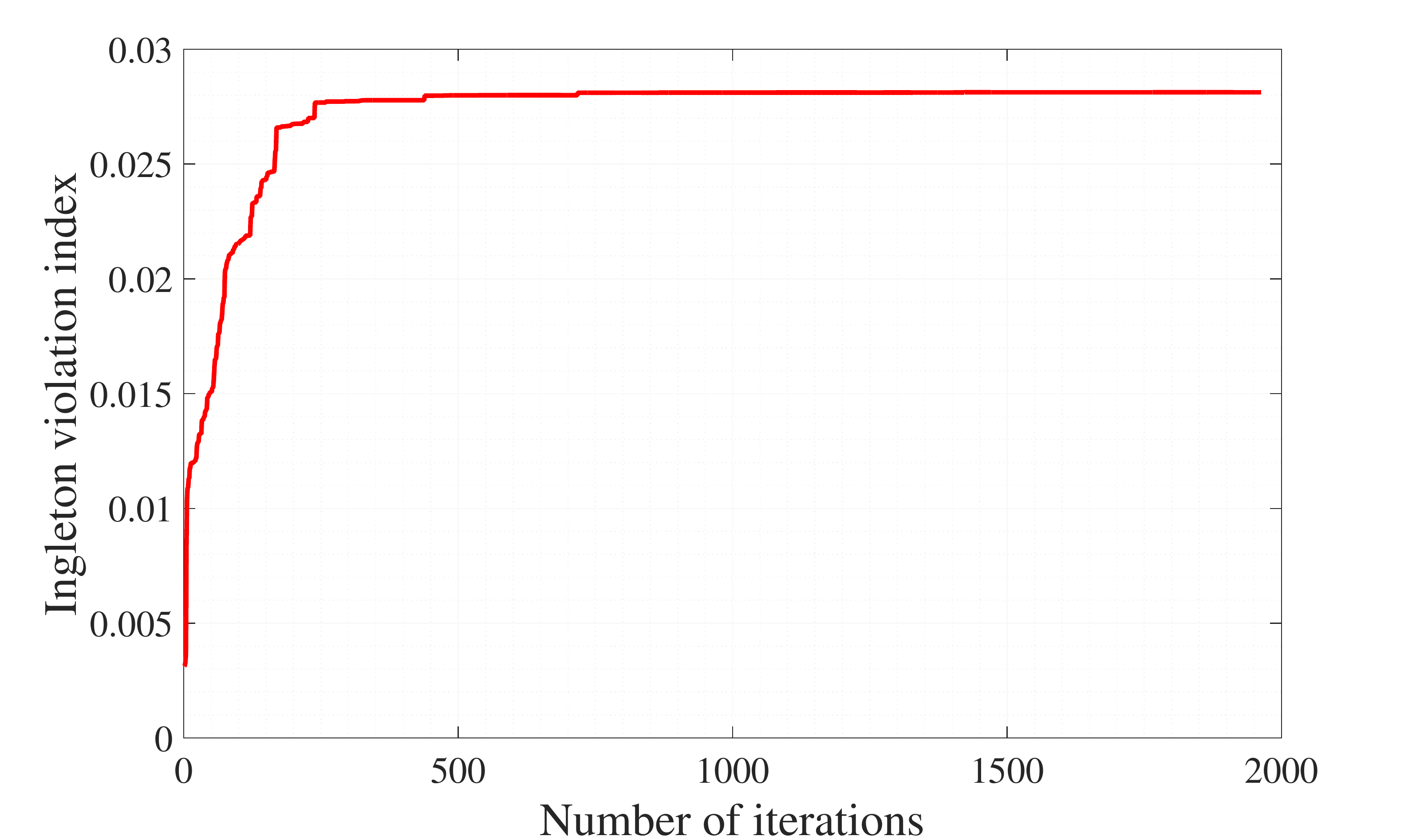}}
 	\caption{~\subref{IngAndViolationIndexVsIteration} The Ingleton violation indices  and corresponding  Ingleton scores for the $100$ points. ~\subref{OurViolationIndexFor2As}  The convergence of the maximum violation index.}
 	\label{ConvgIngAndViolationIndexVsIteration}
 \end{figure}

\textbf{Maximization of $\iota(\mathbf{h})$}: In the algorithm, we have tried different initial distributions. Now we discuss the initial distribution that has delivered the best Ingleton index so far. The initial distribution is the distribution of a point near to the centroid ray of the extreme rays obtained by the intersection of $\Gamma_4$ and hyperplanes $I(X_3;X_4)=0, I(X_3;X_4|X_1,X_2)=0$. Our motivation for trying such an initial point is that the points attaining infimum $\mathbb{I}$ lie at the intersection of these two hyperplanes \cite{MatCsi16}. 
We have maximized violation index. The maximum violation index $\iota(\mathbf{h}) = 0.028131$, is attained by the following joint distribution: 
\begin{equation*}
\begin{split}
&p_{1000}= 0.344902567013607, p_{0010} = 0.155146574443413,
p_{0001}=0.155179836273433, p_{0101}=0.000000220372567, \\&p_{1001}=0.000000000059633, p_{0011}=0.000000000000338, p_{0111}=0.344770801837009,
\end{split}
\end{equation*}
and else zero. We have run the algorithm $100$ times to maximize violation index. In Figure~\ref*{ConvgIngAndViolationIndexVsIteration}~\subref{IngAndViolationIndexVsIteration}, violation indices  and corresponding  Ingleton scores for the $100$ points are plotted and it is clear that every time, the algorithm returns index value approximately $0.02813$.  In Figure~\ref*{ConvgIngAndViolationIndexVsIteration}~\subref{OurViolationIndexFor2As}, we have shown the convergence corresponding to the maximum violation index. It is clear that the achieved value is better with less number of iterations ($1963$)  than the value and number of iterations given in \cite{Sha11}.
 
\textbf{Minimization of} $\mathbb{I}(t(\mathbf{h}^\text{ti}))$: 
	The minimum Ingleton scores of $\mathbf{h}^\text{ti}$ and of $t(\mathbf{h}^\text{ti})$ for four random variables with alphabet size $5^4$, are found to be $-0.091287$ and $-0.092499$ approximately. Following is the procedure by which we have obtained both the scores and the corresponding distributions. 
	\begin{itemize}
		\item We determined hyperplanes of the extreme rays of the intersection of base points of the pyramid and the tight polymatroid of $\mathbf{h}$ given in the \cite[Example 2]{MatCsi16}. 
		\item Run the algorithm taking $\mathbf{h}^\text{ti}$ as a target and the initial distribution as uniform distribution for four random variables with alphabet $\{0,1,2,3,4\}$ each. We reduced normalized distance and alternatively $14$ hyperlane score reduction (described below), then we found the distribution (say $P$) which gives the nearest point (say $f$) to $\mathbf{h}^\text{ti}$.
		\item Use $P$ as the initial distribution in the algorithm to minimize $\mathbb{I}(t(\mathbf{h}^{\text{ti}}))$. In this way we found the infimal Ingleton scores.
		\end{itemize}
	The Ingleton score  $ \mathbb{I}(\mathbf{h}) = -0.0650661$ is obtained by the joint distribution for four random variables with alphabet $\{0,1,2,3,4\}$ given in Section \ref{sec:app1}.
We note that, for this point $\mathbf{h}$, the Ingleton score of the tight component $\mathbb{I}(\mathbf{h}^\text{ti}) = -0.091287$ and the Ingleton score of its linear transformation $\mathbb{I}(t(\mathbf{h}^\text{ti})) = -0.092499$, are smaller (better) than the scores given in \cite{MatCsi16}.

\subsubsection{A New Approach for Normalized Distance Reduction}\label{sec:2b}
To restrict the search area, we propose incorporating
alternative hyperplane score reduction in addition to
normalized distance reduction. The basic idea for this new
approach is the observation that a ray in $2^n-1$ dimensional
Euclidean space can be described by the intersection of $2^n-2$ independent hyperplanes through the origin. For a given point
in the pyramid, we obtain hyperplane representation from the
point and $14$ base points. This gives $15$ hyperplanes including
reverse Ingleton hyperplane. For $14$ non-Ingeton hyperplanes
we define hyperplane scores.

\begin{definition}[Hyperplane through origin]
A hyperplane through origin can be described by the coefficient vector $$\mathbf{g}=[g_{\alpha}:\alpha \subseteq \N\setminus\emptyset]$$ and the corresponding hyperplane is
$$H_\mathbf{g}=\left\{\mathbf{h}: \sum_{\alpha} g_{\alpha}h_{\alpha}=0\right\}.$$
 \end{definition}

\begin{definition}[Hyperplane score]
Consider a point  $\mathbf{h}$ and a hyperplane $H_\mathbf{g}$. The hyperplane score $s_{H_{\mathbf{g}}}$ for $\mathbf{h}$ is defined as 
$$s_{H_{\mathbf{g}}}(\mathbf{h})= \frac{\sum_{\alpha}g_\alpha h_\alpha}{h_\N}.$$
 \end{definition}
\begin{example}[Hyperplane scores]
Consider the point (FC) obtained by the distribution given in \cite{DouFreZeg11}: 
\begin{equation*}
\begin{split}
&\mathbf{h}=  [0.9345\;0.9345\;1.5811\;1.0000\;1.4401\;1.4401\;1.8802 \;1.0000\;1.4401\;1.4401\;1.8802\;1.8802\;1.8802\;1.8802\;1.8802]
\end{split}
\end{equation*}
We obtain the hyperplane representation, given in Table \ref{HyperplanesFCandBP} (excluding Ingleton hyperplane), from the extreme ray representation of FC and $14$ base points of the pyramid.	

\end{example} 
\begin{table}[htp!]
\scriptsize
		\centering
		\caption{The coefficients of the hyperplanes (excluding Ingleton) defined by FC and $14$ base points of the pyramid.}
		\label{HyperplanesFCandBP}
		\begin{tabular}{|c|ccccccccccccccc|}
			\hline
			No. & $g_1$ & $g_2$ & $g_{12}$ & $g_3$ & $g_{13}$ & $g_{23}$ & $g_{123}$ & $g_{4}$ & $g_{14}$ & $g_{24}$  & $g_{124}$ & $g_{34}$ & $g_{134}$ & $g_{234}$ & $g_{1234}$\\ \hline 
			1 & 0& 0& 0& 0& 0& 0& 0& 0& 0& 0& 0& -1.0000 &  1.0000 &  1.0000 & -1.0000 \\ \hline 
			2 & -1.0000& -3.5645&  1.0000& 0&  1.0000&  3.5645& -1.0000& 0&  1.0000&  3.5645& -1.0000& -1.0000& 0& -2.5645& 0 \\ \hline 
			3 & 0& 0& 0& 0& 0& 0& 0& 0& 0& 0& -1.0000& 0& 0& 0&  1.0000 \\ \hline 
			4 & -1.0000& -1.8136&  1.8136& 0&  1.0000&  1.8136& -1.8136& 0&  1.0000&  1.0000& -1.0000& -1.0000& 0& 0& 0 \\ \hline 
			5 & -1.8136& -1.0000&  1.8136& 0&  1.8136&  1.0000& -1.8136& 0&  1.0000&  1.0000& -1.0000& -1.0000& 0& 0& 0 \\ \hline 
			6 & 0& 0& 0& -1.0000&  1.0000&  1.0000& -1.0000& 0& 0& 0& 0& 0& 0& 0& 0 \\ \hline 
			7 & 0& 0& 0& 0& 0& 0& -1.0000& 0& 0& 0& 0& 0& 0& 0&  1.0000 \\ \hline 
			8 & -1.0000& -1.8136&  1.8136& 0&  1.0000&  1.0000& -1.0000& 0&  1.0000&  1.8136& -1.8136& -1.0000& 0& 0& 0 \\ \hline 
			9 & -1.8136& -1.0000&  1.8136& 0&  1.0000&  1.0000& -1.0000& 0&  1.8136&  1.0000& -1.8136& -1.0000& 0& 0& 0 \\ \hline 
			10 & 0& 0& 0& 0& 0& 0& 0& -1.0000&  1.0000&  1.0000& -1.0000& 0& 0& 0& 0 \\ \hline 
			11 & -1.0000& -1.0000&  1.0000&  1.4023&  1.0000&  1.0000& -1.0000&  1.4023&  1.0000&  1.0000& -1.0000& -2.4023& 0& 0& 0 \\ \hline 
			12 & -3.5645& -1.0000&  1.0000& 0&  3.5645&  1.0000& -1.0000& 0&  3.5645&  1.0000& -1.0000& -1.0000& -2.5645& 0& 0 \\ \hline 
			13 & 0& 0& 0& 0& 0& 0& 0& 0& 0& 0& 0& 0& -1.0000& 0&  1.0000 \\ \hline 
			14 & 0& 0& 0& 0& 0& 0& 0& 0& 0& 0& 0& 0& 0& -1.0000&  1.0000 \\ \hline 
		\end{tabular}
	\end{table}

To obtain the nearest point to $\mathbf{h}$, first, find the centroid ray $E_{\mathbf{c}}$ of base points of the pyramid. Use the algorithm to find a point $\mathbf{c}'$ approximately in $E_{\mathbf{c}}$ and uniform distribution as the initial distribution for four binary random variables. Let the nearest point be $\mathbf{c}'$ and its corresponding distribution be $\mathbf{p}'$.

\begin{case}\label{FCbasePointDistance}
 In the algorithm, start with the parameters $\mathbf{h}_t = \mathbf{h}$ and $\mathbf{p}_s =  \mathbf{p}'$ to find nearest point to $\mathbf{h}$.  
We have obtained $100$ points and their corresponding distributions. Of them, the nearest to $\mathbf{h}$ found is
\begin{equation*}
\begin{split}
&[0.9345\;0.9345\;1.5811\;1.0000\;1.4401\;1.4401\;1.8802\;1.0000\;1.4401\;1.4401\;1.8802\;1.8801\;1.8802\;1.8802\;1.8802]
\end{split}
\end{equation*}
with Ingleton score $ -0.08936339$ and is obtained by the following joint distribution:
\begin{equation*}
\begin{split}
p_{1100}=0.350471340516428, p_{0110} = 0.149511435377891, p_{0001}=0.000001680821054, p_{0101}=0.149541007916755,\\p_{1001}=0.000000733705598, p_{1101}=0.000001689807316, p_{0011}= 0.350472111852016, p_{0111}=0.000000000002942,
\end{split}
\end{equation*}
and else zero.
\end{case}
\begin{figure}[htp!]
	\centering
	\subfigure[\label{normalizedDistance}]{\includegraphics[scale=0.27]{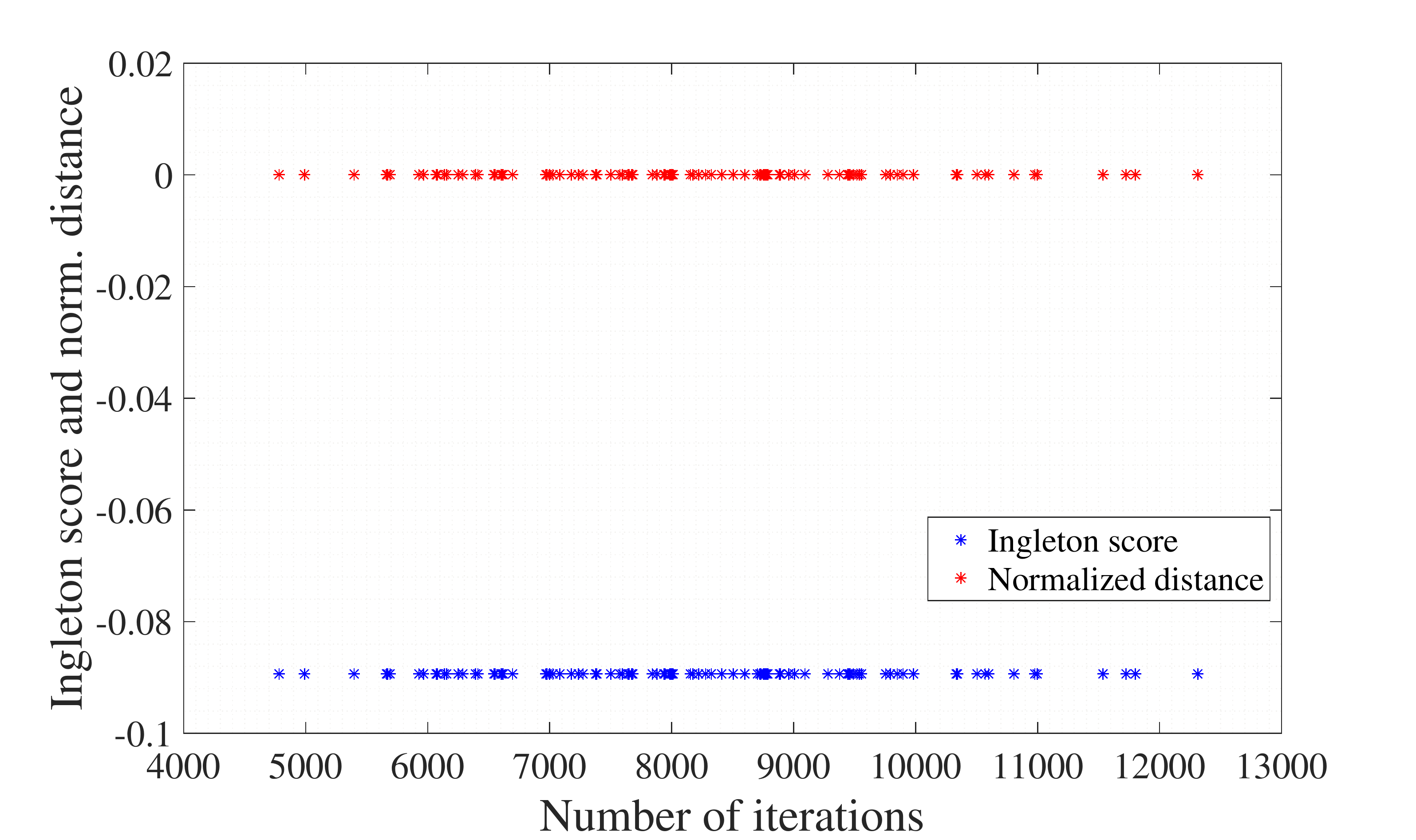}}
	\subfigure[\label{normalizedDistanceHyperplane}]{\includegraphics[scale=0.27]{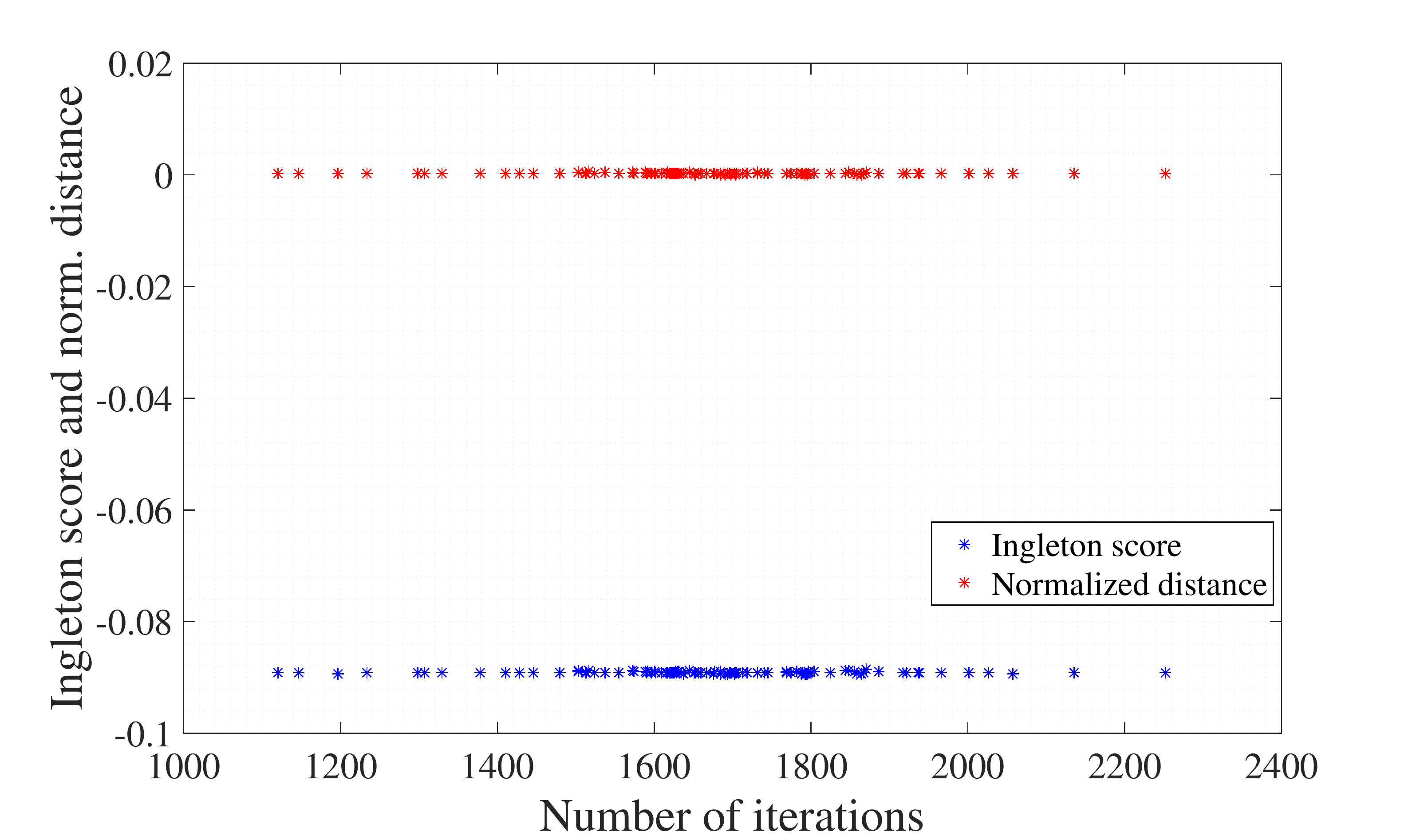}}
	\caption{~\subref{normalizedDistance} Ingleton  scores and distances for Case \ref*{FCbasePointDistance}, ~\subref{normalizedDistanceHyperplane} Ingleton scores and distances for Case \ref*{FCbasePointDistanceHp}.}
	\label{FCIngAndDistance}
\end{figure}
\begin{case}\label{FCbasePointDistanceHp}
We have obtained $14$ hyperplanes by the intersection of $\mathbf{h}$ and $14$ base points of the pyramid. The algorithm is modified as follows: At each iteration a new distribution is accepted such that the normalized distance and one of the fixed hyperplane score is reduced. Once the hyperplane score reaches near to 0 then the algorithm uniformly randomly selects any other hyperplan and continues normalized distance and hyperplane score reduction. In this way, we obtained $100$ points and corresponding distributions. Of them, the nearest to $\mathbf{h}$ is
\begin{equation*}
\begin{split}
&[0.9346\;0.9346\;1.5811\;1.0000\;1.4401\;1.4401\;1.8802\;1.0000\;1.4401\;1.4401\;1.8802\;1.8799\;1.8802\;1.8802\;1.8804]
\end{split}
\end{equation*}
  with Ingeton score $ -0.089320$ and is obtained by the following joint distribution:
 \begin{equation*}
 \begin{split}
 p_{0000}=0.000000000000033, p_{1000}= 0.000000000002665, p_{1100}=0.350535077295177, p_{0110} = 0.149338731893951,\\ p_{0001}= 0.000014509393386,  p_{0101}=0.149518932564159, p_{1001}=0.000003672853305, p_{1101}=0.000015752047679,\\ p_{0011}= 0.350573321610050,  p_{0111}=0.000000000000012, p_{1011}=0.000000002265614, p_{1111}=0.000000000073969,
 \end{split}
 \end{equation*} 
 and else zero.
\end{case}
Ingleton scores and distances for $100$ points of the above two cases is plotted in Figure~\ref*{FCIngAndDistance}. Note that the convergence in Case \ref*{FCbasePointDistanceHp} takes less number of iterations than in Case \ref*{FCbasePointDistance}. For these two cases all returned points are very near to the target. 

\begin{figure}[htp!]
	\centering
	\subfigure[\label{normalizedDistanceUni}]{\includegraphics[scale=0.27]{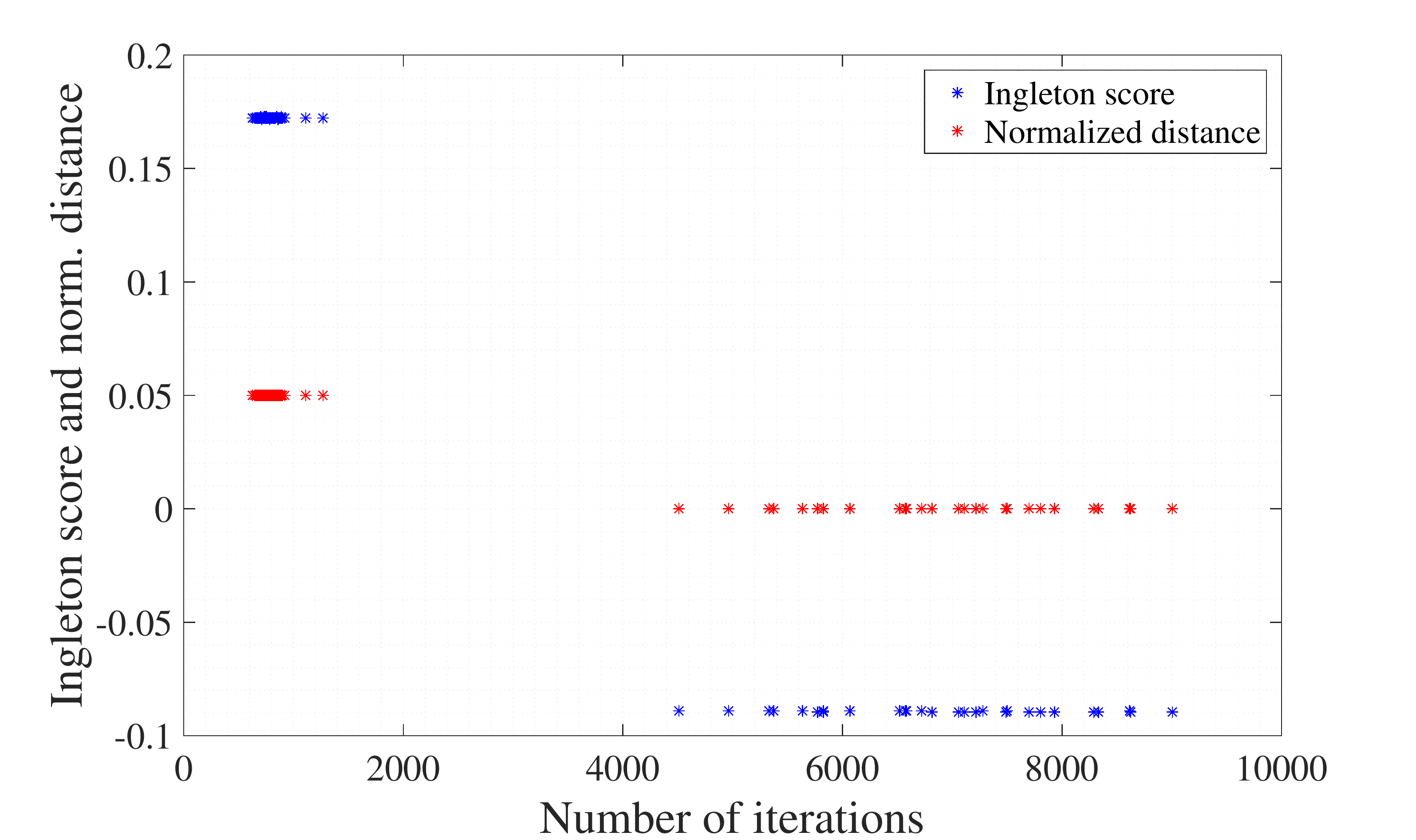}}
	\subfigure[\label{normalizedDistanceHyperplaneUni}]{\includegraphics[scale=0.27]{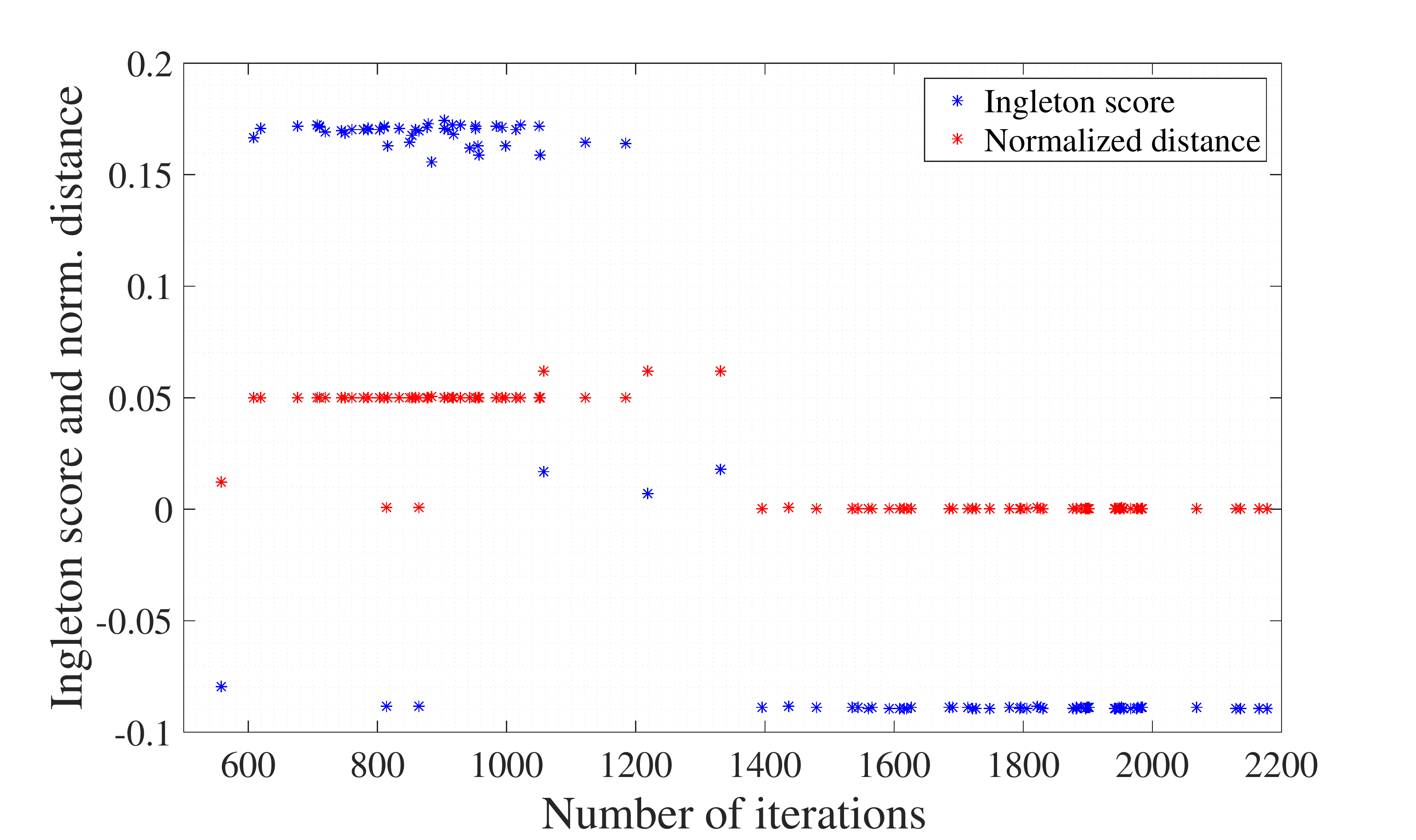}}
	\caption{~\subref{normalizedDistanceUni} Ingleton scores and normalized distances for Case \ref*{FCbasePointDistance} with the initial distribution as uniform distribution, ~\subref{normalizedDistanceHyperplaneUni} Ingleton scores and normalized distances for Case \ref*{FCbasePointDistanceHp} with the initial distribution as uniform distribution.}
	\label{FCIngAndDistanceUni}
\end{figure}

We have repeated the above two cases with initial distribution as uniform distribution for binary random variables. The numerical result is shown in Figure \ref*{FCIngAndDistanceUni} for $100$ points.

From Figure~\ref*{FCIngAndDistanceUni}, It is clear that there are points returned by the algorithm which are not near to the target. For example, the point
$$[0.8823 \;0.9599 \;1.6839 \;0.9013 \;1.5778 \;1.5772 \;1.7887 \;0.9962 \;1.5265 \;1.4657 \;1.9796 \;1.8679 \;1.9710 \;1.9731 \;2.0745]$$
with distance $0.049937588319301$ from the target,  and Ingleton score $0.170042927792174$, is not converging but one of its $14$ hyperplane scores
$$[0.0017 \;0.6347 \;0.0949 \;0.7697 \;0.8334 \;0.4650 \;0.2858 \;0.5237 \;0.6363 \;0.0164 \;0.3944 \;0.9966 \;0.1035 \;0.1014].$$
is close to zero and hence the point is near to the boundary of $\overline\Gamma_4^*$.

\subsubsection{Inner Bounds from known Outer Bounds}\label{sec:2c}
As discussed in Section \ref{sec: Algorithm Extensions and Comparison}, the characterization of $\overline{\Gamma}^*_4$ reduces to the entropic points in this pyramid defined by 15 extreme rays \cite{DouFreZeg11} (actually, the characterization is further reduced to the 11-dimensional sub-cone of tight points \cite{MatCsi16}) and hence we focus on the entropic region in the pyramid. The pyramid is further cut-off by two ZY98 \cite{ZY98} inequalities. This removes one extreme ray (associated with V\'{a}mos polymatroid) and introduces 9 additional extreme rays. This outer bound with 23 extreme rays (or equivalently, 17 hyperplanes) has a relative volume of 98.4568\% \cite{DouFreZeg11}. Our inner bound with the same description complexity has relative volume 43.8026\%. The nine extreme rays, $E_{i}, i=1,\ldots,9$, defining the inner bound togather with the base extreme rays are listed in Table \ref{tab:3}. The volume of the inner bound obtained by 14 based points and four-atom conjecture point is 35.7493\%.
 \footnote{It appears that the procedure to compute the volumes in \cite[Table 4.3]{Liu16} may be different or there may be a numerical error in the computed volume 43.5\%. We have verified that our process to compute the volume is the same as that in \cite{DouFreZeg11} by computing volume of ZY98 outer bound in the paper.}
  Moreover, including the four-atom conjecture point the relative volume of resulting inner bound is 46.6521\%. 
\begin{table}
\scriptsize
\centering
\caption{The 9 nearest points $E_i$ corresponding to the 9 extreme rays $Z_i, i= 1,\dots, 9.$}
\label{tab:3}
\begin{tabular}{|c|ccccccccccccccc|}
\hline
No. & $h_1$ & $h_2$ & $h_{12}$ & $h_3$ & $h_{13}$ & $h_{23}$ & $h_{123}$ & $h_{4}$ & $h_{14}$ & $h_{24}$  & $h_{124}$ & $h_{34}$ & $h_{134}$ & $h_{234}$ & $h_{1234}$ \\ \hline  
$Z_1$ & 0.3333  & 0.6667  & 0.8333  & 0.5000  & 0.6667  & 0.8333  & 1.0000  & 0.5000  & 0.6667  & 0.8333  & 1.0000  & 1.0000  & 1.0000  & 1.0000  & 1.0000 \\ \hline
$E_1$ &0.4364 &0.8716 &1.1321 &0.6992 &0.9088 &1.1454 &1.3488 &0.6992 &0.9087 &1.1454 &1.3488 &1.3286 &1.3540 &1.3467 &1.3560 \\ \hline
$Z_2$ & 0.5000  & 0.5000  & 0.8333  & 0.5000  & 0.8333  & 0.6667  & 1.0000  & 0.5000  & 0.6667  & 0.8333  & 1.0000  & 1.0000  & 1.0000  & 1.0000  & 1.0000 \\ \hline
$E_2$ & 0.9114 &0.9108 &1.5851 &0.9747 &1.5836 &1.2700 &1.8740 &0.9750 &1.2702 &1.5837 &1.8740 &1.8603 &1.8786 &1.8786 &1.8920 \\ \hline
$Z_3$ & 0.6667  & 0.5000  & 0.8333  & 0.3333  & 0.8333  & 0.6667  & 1.0000  & 0.6667  & 0.8333  & 0.8333  & 1.0000  & 1.0000  & 1.0000  & 1.0000  & 1.0000 \\ \hline
$E_3$ &0.5963 &0.4156 &0.7384 &0.3208 &0.7371 &0.5908 &0.8651 &0.5897 &0.7307 &0.7285 &0.8567 &0.8102 &0.8546 &0.9026 &0.9398 \\ \hline
$Z_4$ & 0.6667  & 0.3333  & 0.8333  & 0.5000  & 0.8333  & 0.6667  & 1.0000  & 0.5000  & 0.8333  & 0.6667  & 1.0000  & 1.0000  & 1.0000  & 1.0000  & 1.0000 \\ \hline
$E_4$ & 0.8837 &0.4445 &1.1525 &0.7121 &1.1668 &0.9257 &1.3752 &0.7126 &1.1666 &0.9260 &1.3750 &1.3567 &1.3733 &1.3791 &1.3818 \\ \hline
$Z_5$ & 0.5000  & 0.6667  & 0.8333  & 0.3333  & 0.6667  & 0.8333  & 1.0000  & 0.6667  & 0.8333  & 0.8333  & 1.0000  & 1.0000  & 1.0000  & 1.0000  & 1.0000 \\ \hline
$E_5$ & 0.4050 &0.5819 &0.7184 &0.3114 &0.5752 &0.7172 &0.8407 &0.5751 &0.7084 &0.7122 &0.8330 &0.7883 &0.8775 &0.8311 &0.9133 \\ \hline
$Z_6$ & 0.5000  & 0.5000  & 0.8333  & 0.5000  & 0.6667  & 0.8333  & 1.0000  & 0.5000  & 0.8333  & 0.6667  & 1.0000  & 1.0000  & 1.0000  & 1.0000  & 1.0000 \\ \hline
$E_6$ & 0.9103 &0.9103 &1.5847 &0.9743 &1.2692 &1.5832 &1.8734 &0.9743 &1.5832 &1.2692 &1.8734 &1.8597 &1.8779 &1.8779 &1.8901 \\ \hline
$Z_7$ & 0.6667  & 0.5000  & 0.8333  & 0.6667  & 0.8333  & 0.8333  & 1.0000  & 0.3333  & 0.8333  & 0.6667  & 1.0000  & 1.0000  & 1.0000  & 1.0000  & 1.0000 \\ \hline
$E_7$ & 0.5875 &0.4096 &0.7261 &0.5811 &0.7194 &0.7165 &0.8422 &0.3149 &0.7249 &0.5815 &0.8500 &0.7967 &0.8402 &0.8873 &0.9237 \\ \hline
$Z_8$ & 0.5000  & 0.6667  & 0.8333  & 0.6667  & 0.8333  & 0.8333  & 1.0000  & 0.3333  & 0.6667  & 0.8333  & 1.0000  & 1.0000  & 1.0000  & 1.0000  & 1.0000 \\ \hline
$E_8$  & 0.4190 &0.6005 &0.7442 &0.5942 &0.7347 &0.7361 &0.8636 &0.3236 &0.5955 &0.7429 &0.8721 &0.8165 &0.9099 &0.8616 &0.9477 \\ \hline
$Z_9$ & 0.6667  & 0.6667  & 0.8333  & 0.5000  & 0.8333  & 0.8333  & 1.0000  & 0.5000  & 0.8333  & 0.8333  & 1.0000  & 1.0000  & 1.0000  & 1.0000  & 1.0000 \\ \hline
$E_9$ & 0.7030 &0.7031 &0.8352 &0.5180 &0.8872 &0.8874 &1.0055 &0.5180 &0.8872 &0.8874 &1.0055 &0.9675 &1.0439 &1.0442 &1.1205 \\ \hline
\end{tabular}
\end{table}

\subsubsection{Inner Bounds using a Grid Based Approach}\label{sec:2d}
Now we present a new and simple grid based approach to obtain inner bounds for the entropy region within a cone defined by some collection of extreme rays $E_i$'s. 
\begin{definition}
A grid $G$, for a given collection of rays $E_i$'s and for given set of possible values of $\lambda_i$ in $\mathbb{R}^+$, is the set of rays
$${E}_g= \sum_{i} \lambda_i {E}_i.$$
\end{definition}
Note that the grid rays (or points) are obtained by conic combination of the extreme rays $E_i$'s, and hence the grid elements (rays) are in the conic hull of $E_i$'s. 

For four random variable case, the pyramid is defined by the ``V\'{a}mos'' extreme ray ${E_v}$ (containing a V\'{a}mos polymatroid) and other 14 base extreme rays ${E_b}, b\in B$. 
If we consider $\lambda_i \in \{0,a\}$ for some $a>0$ such that $\sum_i \lambda_i=a$ for extreme rays of the pyramid then we get the extreme rays defining the pyramid itself. If we increase the range of $\lambda_i$ we can obtain finer grids with more rays. We consider the grids defined as follows.

\begin{align*}
G_1&\triangleq \left\{E_v, E_b: b\in B\right\}\\
G_2&\triangleq \left\{E: E=E_v+E_b, b\in B\right\}\\
G_3&\triangleq \left\{E: E=E_v+E_b+E_c \text{ for distinct }b,c \in B\right\}\\
G_4&\triangleq \left\{E: E=E_v+E_b+E_c+E_d \text{ for distinct } b,c,d\in B\right\}\\
G_5&\triangleq \left\{E: E=E_v+E_b+E_c+E_d+E_e  \text{ for distinct } b,c,d,e \in B \right\}
\end{align*}

For the grids $G_1, G_2, G_3, G_4, G_5$ so defined, here is the summary of simulation results:
\begin{itemize}
\item $G_1$ has 15 elements of which only 1 violates Ingleton inequality and corresponding nearest distribution known so far is that of 4-atom conjecture point. The inner bound has relative volume 35.7493\% (as discussed in Section \ref{sec:2c}).  
\item $G_2$ has 14 elements (rays). We obtained nearest points of which only 10 violate Ingleton inequality. Thus we obtain inner bound using $G_1 \cup G_2$ with 25 extreme rays and relative volume \textbf{50.0869\%}.
\item $G_3$ has 91 elements. We obtained nearest points of which only 31 violate Ingleton inequality. Thus we obtain inner bound using $G_1 \cup G_3$ with 46 extreme rays and relative volume 47.3243\%. We also obtain inner bound using $G_1 \cup G_2 \cup G_3$ with 56 extreme rays and relative volume \textbf{51.6197\%}.
\item $G_4$ has 364 elements. We obtained nearest points of which only 54 violate Ingleton inequality. Thus we obtain inner bound using $G_1 \cup G_4$ with 69 extreme rays and relative volume 47.8652\%. We also obtain inner bound using $G_1 \cup G_2 \cup G_3\cup G_4$ with 110 extreme rays and relative volume \textbf{53.3164\%}.
\item $G_5$ has 1001 elements. We obtained nearest points of which only 175 violate Ingleton inequality. Thus we obtain inner bound using $G_1 \cup G_5$ with 190 extreme rays and relative volume 52.2465\%. We also obtain inner bound using $G_1 \cup G_2 \cup G_3\cup G_4 \cup G_5$ with 285 extreme rays and relative volume \textbf{56.2590\%}\footnote{This is better than the relative volume 53.4815\% reported in \cite{DouFreZeg11}. More importantly, we are interested in inner bounds with minimum possible ``description complexity'' and in \cite{DouFreZeg11} the number of extreme rays is not specified where as in \cite{Liu16} the extreme rays used are more than 1000 to obtain an inner bound a relative volume 57.8\%. However, as mentioned in footnote 4, the procedure to compute the volumes in [16, Table 4.3] may be different or there may be a numerical error.}. This is the best inner bound for the entropy region in the pyramid so far in terms of relative volume with this (minimum) number of extreme rays.
\end{itemize}

\subsubsection{Points near to Ingleton Cone Extreme Rays}\label{sec:2e}
\begin{definition}[Ingleton cone]
	The Ingleton cone for four random variables is defined by the $35$ extreme rays $\rho_{i}, i=1,\ldots,35$ given in Table \ref{IngletonConeExtremerays} (see also \cite[Figure 5]{Hammer00}). These $35$ extreme rays are obtained by the intersection of $\Gamma_4$ and six Ingleton inequalities.
\end{definition}
Using the algorithm, we have computed points $\mathbf{h}_i$ near to the extreme rays. The total of 26 points thus obtained are very close to the respective extreme rays and hence approximately binary entropic as it can be confirmed from \cite{WalWeb11}. For the remaining 9 extreme rays (the corresponding rows in Table  \ref{IngletonConeExtremerays} are shaded gray), near binary entropic points are also found. From the results in \cite{WalWeb11},  we verify that all these 9 points are not binary entropic. In particular, \cite[Table 1]{WalWeb11} lists all (up to permutation) non-binary entropic points of normalized entropic region. Note that, $\rho_{26}$ is the 5th entry in \cite[Table 1]{WalWeb11}, $\rho_{28}$-$\rho_{31}$ are (up to permutation) the 1th entry in \cite[Table 1]{WalWeb11} and $\rho_{32}$-$\rho_{35}$ are (up to permutation) the 8th entry in \cite[Table 1]{WalWeb11}.

\definecolor{Gray}{gray}{0.9}
\begin{table}[H]
\scriptsize
	\centering
	\caption{Extreme rays $\rho_i$ of the Ingleton cone and corresponding near binary entropic points $\mathbf{h}_i$.}
	\label{IngletonConeExtremerays}
	\begin{tabular}{|c|ccccccccccccccc|}
		\hline
		{ } & $h_1$ & $h_2$ & $h_{12}$ & $h_3$ & $h_{13}$ & $h_{23}$ & $h_{123}$ & $h_{4}$ & $h_{14}$ & $h_{24}$ & $h_{124}$ & $h_{34}$ & $h_{134}$ & $h_{234}$ & $h_{1234}$
		 \\ \hline 
		 $\rho_{1}$ & 1 & 1 & 1 & 1 & 1 & 1 & 1 & 1 & 1 & 1 & 1 & 1 & 1 & 1 & 1 \\ \hline
			$\mathbf{h}_1$ & 1.0000 & 1.0000 & 1.0002 & 1.0000 & 1.0002 & 1.0002 & 1.0003 & 1.0000 & 1.0002 & 1.0002 & 1.0003 & 1.0002 & 1.0003 & 1.0003 & 1.0003  \\ \hline
		$\rho_{2}$ & 1 & 1 & 1 & 1 & 1 & 1 & 1 & 0 & 1 & 1 & 1 & 1 & 1 & 1 & 1 \\ \hline
			$\mathbf{h}_2$ & 0.9998 & 0.9998 & 1.0000 & 0.9998 & 1.0000 & 1.0000 & 1.0001 & 0.0001 & 1.0000 & 1.0000 & 1.0001 & 1.0000 & 1.0002 & 1.0002 & 1.0003  \\ \hline
		$\rho_{3}$ & 1 & 1 & 1 & 0 & 1 & 1 & 1 & 1 & 1 & 1 & 1 & 1 & 1 & 1 & 1 \\ \hline
			$\mathbf{h}_3$ & 0.9999 & 0.9999 & 1.0001 & 0.0001 & 1.0000 & 1.0000 & 1.0002 & 0.9999 & 1.0001 & 1.0001 & 1.0002 & 1.0000 & 1.0002 & 1.0002 & 1.0003  \\ \hline
		$\rho_{4}$ & 1 & 0 & 1 & 1 & 1 & 1 & 1 & 1 & 1 & 1 & 1 & 1 & 1 & 1 & 1 \\ \hline
			$\mathbf{h}_{4}$ & 0.9998 & 0.0002 & 1.0000 & 0.9998 & 1.0000 & 1.0000 & 1.0002 & 0.9998 & 1.0000 & 1.0000 & 1.0002 & 1.0000 & 1.0002 & 1.0002 & 1.0003  \\ \hline			$\rho_{5}$ & 0 & 1 & 1 & 1 & 1 & 1 & 1 & 1 & 1 & 1 & 1 & 1 & 1 & 1 & 1 \\ \hline
			$\mathbf{h}_{5}$ & 0.0002 & 0.9999 & 1.0000 & 0.9999 & 1.0000 & 1.0001 & 1.0002 & 0.9999 & 1.0000 & 1.0001 & 1.0002 & 1.0001 & 1.0002 & 1.0002 & 1.0004  \\ \hline			$\rho_{6}$ & 1 & 1 & 1 & 0 & 1 & 1 & 1 & 0 & 1 & 1 & 1 & 0 & 1 & 1 & 1 \\ \hline 
			$\mathbf{h}_6$ & 0.9999 & 0.9999 & 1.0000 & 0.0001 & 1.0000 & 1.0000 & 1.0001 & 0.0001 & 1.0000 & 1.0000 & 1.0001 & 0.0001 & 1.0000 & 1.0000 & 1.0001  \\ \hline				$\rho_{7}$ & 1 & 0 & 1 & 1 & 1 & 1 & 1 & 0 & 1 & 0 & 1 & 1 & 1 & 1 & 1 \\ \hline
			$\mathbf{h}_{7}$ & 0.9999 & 0.0000 & 1.0000 & 0.9999 & 1.0000 & 1.0000 & 1.0001 & 0.0000 & 1.0000 & 0.0001 & 1.0000 & 1.0000 & 1.0001 & 1.0000 & 1.0001  \\ \hline	
		$\rho_{8}$ & 1 & 0 & 1 & 0 & 1 & 0 & 1 & 1 & 1 & 1 & 1 & 1 & 1 & 1 & 1 \\ \hline
			$\mathbf{h}_{8}$ & 0.9999 & 0.0000 & 1.0000 & 0.0000 & 1.0000 & 0.0001 & 1.0000 & 0.9999 & 1.0000 & 1.0000 & 1.0000 & 1.0000 & 1.0000 & 1.0000 & 1.0001  \\ \hline
		$\rho_{9}$ & 0 & 1 & 1 & 1 & 1 & 1 & 1 & 0 & 0 & 1 & 1 & 1 & 1 & 1 & 1 \\ \hline
			$\mathbf{h}_{9}$ & 0.0001 & 0.9999 & 1.0000 & 0.9999 & 1.0000 & 1.0001 & 1.0001 & 0.0001 & 0.0001 & 1.0000 & 1.0000 & 1.0000 & 1.0000 & 1.0001 & 1.0002  \\ \hline			$\rho_{10}$ & 0 & 1 & 1 & 0 & 0 & 1 & 1 & 1 & 1 & 1 & 1 & 1 & 1 & 1 & 1 \\ \hline
			$\mathbf{h}_{10}$ & 0.0000 & 1.0000 & 1.0000 & 0.0000 & 0.0000 & 1.0000 & 1.0000 & 1.0000 & 1.0000 & 1.0000 & 1.0000 & 1.0000 & 1.0000 & 1.0000 & 1.0001  \\ \hline			$\rho_{11}$ & 0 & 0 & 0 & 1 & 1 & 1 & 1 & 1 & 1 & 1 & 1 & 1 & 1 & 1 & 1 \\ \hline
			$\mathbf{h}_{11}$ & 0.0001 & 0.0001 & 0.0001 & 0.9999 & 1.0000 & 1.0000 & 1.0000 & 0.9999 & 1.0000 & 1.0000 & 1.0000 & 1.0001 & 1.0001 & 1.0001 & 1.0002  \\ \hline
		$\rho_{12}$ & 1 & 0 & 1 & 0 & 1 & 0 & 1 & 0 & 1 & 0 & 1 & 0 & 1 & 0 & 1 \\ \hline
			$\mathbf{h}_{12}$ & 0.9999 & 0.0001 & 1.0000 & 0.0001 & 1.0000 & 0.0002 & 1.0001 & 0.0001 & 1.0000 & 0.0002 & 1.0001 & 0.0002 & 1.0001 & 0.0003 & 1.0002  \\ \hline 		$\rho_{13}$ & 0 & 1 & 1 & 0 & 0 & 1 & 1 & 0 & 0 & 1 & 1 & 0 & 0 & 1 & 1 \\ \hline
			$\mathbf{h}_{13}$ & 0.0001 & 0.9999 & 1.0000 & 0.0001 & 0.0003 & 1.0000 & 1.0002 & 0.0001 & 0.0003 & 1.0000 & 1.0002 & 0.0003 & 0.0004 & 1.0002 & 1.0003  \\ \hline			$\rho_{14}$ & 0 & 0 & 0 & 1 & 1 & 1 & 1 & 0 & 0 & 0 & 0 & 1 & 1 & 1 & 1 \\ \hline
			$\mathbf{h}_{14}$ & 0.0001 & 0.0001 & 0.0002 & 0.9999 & 1.0000 & 1.0000 & 1.0001 & 0.0001 & 0.0002 & 0.0002 & 0.0003 & 1.0000 & 1.0001 & 1.0001 & 1.0002  \\ \hline
		$\rho_{15}$ & 0 & 0 & 0 & 0 & 0 & 0 & 0 & 1 & 1 & 1 & 1 & 1 & 1 & 1 & 1 \\ \hline
			$\mathbf{h}_{15}$ & 0.0001 & 0.0001 & 0.0002 & 0.0001 & 0.0002 & 0.0002 & 0.0004 & 0.9999 & 1.0000 & 1.0000 & 1.0001 & 1.0000 & 1.0001 & 1.0001 & 1.0003  \\ \hline
		$\rho_{16}$ & 1 & 1 & 2 & 1 & 2 & 2 & 2 & 0 & 1 & 1 & 2 & 1 & 2 & 2 & 2 \\ \hline
			$\mathbf{h}_{16}$ & 1.0000 & 1.0000 & 2.0000 & 1.0000 & 2.0000 & 2.0000 & 2.0000 & 0.0000 & 1.0000 & 1.0000 & 2.0000 & 1.0000 & 2.0000 & 2.0000 & 2.0000  \\ \hline			$\rho_{17}$ & 1 & 1 & 2 & 0 & 1 & 1 & 2 & 1 & 2 & 2 & 2 & 1 & 2 & 2 & 2 \\ \hline
			$\mathbf{h}_{17}$ & 1.0000 & 1.0000 & 2.0000 & 0.0000 & 1.0000 & 1.0000 & 2.0000 & 1.0000 & 2.0000 & 2.0000 & 2.0000 & 1.0000 & 2.0000 & 2.0000 & 2.0000  \\ \hline			$\rho_{18}$ & 1 & 0 & 1 & 1 & 2 & 1 & 2 & 1 & 2 & 1 & 2 & 2 & 2 & 2 & 2 \\ \hline
			$\mathbf{h}_{18}$ & 1.0000 & 0.0000 & 1.0000 & 1.0000 & 2.0000 & 1.0000 & 2.0000 & 1.0000 & 2.0000 & 1.0000 & 2.0000 & 2.0000 & 2.0000 & 2.0000 & 2.0000  \\ \hline
		$\rho_{19}$ & 0 & 1 & 1 & 1 & 1 & 2 & 2 & 1 & 1 & 2 & 2 & 2 & 2 & 2 & 2 \\ \hline
			$\mathbf{h}_{19}$ & 0.0000 & 1.0000 & 1.0000 & 1.0000 & 1.0000 & 2.0000 & 2.0000 & 1.0000 & 1.0000 & 2.0000 & 2.0000 & 2.0000 & 2.0000 & 2.0000 & 2.0000  \\ \hline			$\rho_{20}$ & 1 & 1 & 2 & 1 & 2 & 2 & 2 & 1 & 2 & 2 & 2 & 1 & 2 & 2 & 2 \\ \hline
			$\mathbf{h}_{20}$ & 1.0000 & 1.0000 & 2.0000 & 1.0000 & 2.0000 & 2.0000 & 2.0000 & 1.0000 & 2.0000 & 2.0000 & 2.0000 & 1.0000 & 2.0000 & 2.0000 & 2.0000  \\ \hline			$\rho_{21}$ & 1 & 1 & 2 & 1 & 2 & 2 & 2 & 1 & 2 & 1 & 2 & 2 & 2 & 2 & 2 \\ \hline
			$\mathbf{h}_{21}$ & 1.0000 & 1.0000 & 2.0000 & 1.0000 & 2.0000 & 2.0000 & 2.0000 & 1.0000 & 2.0000 & 1.0000 & 2.0000 & 2.0000 & 2.0000 & 2.0000 & 2.0000  \\ \hline			$\rho_{22}$ & 1 & 1 & 2 & 1 & 2 & 1 & 2 & 1 & 2 & 2 & 2 & 2 & 2 & 2 & 2 \\ \hline
			$\mathbf{h}_{22}$ & 1.0000 & 1.0000 & 2.0000 & 1.0000 & 2.0000 & 1.0000 & 2.0000 & 1.0000 & 2.0000 & 2.0000 & 2.0000 & 2.0000 & 2.0000 & 2.0000 & 2.0000  \\ \hline
		$\rho_{23}$ & 1 & 1 & 2 & 1 & 2 & 2 & 2 & 1 & 1 & 2 & 2 & 2 & 2 & 2 & 2 \\ \hline
			$\mathbf{h}_{23}$ & 1.0000 & 1.0000 & 2.0000 & 1.0000 & 2.0000 & 2.0000 & 2.0000 & 1.0000 & 1.0000 & 2.0000 & 2.0000 & 2.0000 & 2.0000 & 2.0000 & 2.0000  \\ \hline			$\rho_{24}$ & 1 & 1 & 2 & 1 & 1 & 2 & 2 & 1 & 2 & 2 & 2 & 2 & 2 & 2 & 2 \\ \hline
			$\mathbf{h}_{24}$ & 1.0000 & 1.0000 & 2.0000 & 1.0000 & 1.0000 & 2.0000 & 2.0000 & 1.0000 & 2.0000 & 2.0000 & 2.0000 & 2.0000 & 2.0000 & 2.0000 & 2.0000  \\ \hline			$\rho_{25}$ & 1 & 1 & 1 & 1 & 2 & 2 & 2 & 1 & 2 & 2 & 2 & 2 & 2 & 2 & 2 \\ \hline 
			$\mathbf{h}_{25}$ & 1.0000 & 1.0000 & 1.0000 & 1.0000 & 2.0000 & 2.0000 & 2.0000 & 1.0000 & 2.0000 & 2.0000 & 2.0000 & 2.0000 & 2.0000 & 2.0000 & 2.0000  \\ \hline
		$\rho_{26}$ & 1 & 1 & 2 & 1 & 2 & 2 & 2 & 1 & 2 & 2 & 2 & 2 & 2 & 2 & 2 \\ \hline
		\rowcolor{Gray}
			$\mathbf{h}_{26}$ & 0.7512 & 0.7513 & 1.2950 & 0.7512 & 1.2950 & 1.2950 & 1.4384 & 0.7512 & 1.2950 & 1.2950 & 1.4384 & 1.2950 & 1.4384 & 1.4384 & 1.4384  \\ \hline			$\rho_{27}$ & 1 & 1 & 2 & 1 & 2 & 2 & 3 & 1 & 2 & 2 & 3 & 2 & 3 & 3 & 3 \\ \hline
			$\mathbf{h}_{27}$ & 1.0000 & 1.0000 & 2.0000 & 1.0000 & 2.0000 & 2.0000 & 3.0000 & 1.0000 & 2.0000 & 2.0000 & 3.0000 & 2.0000 & 3.0000 & 3.0000 & 3.0000  \\ \hline			$\rho_{28}$ & 2 & 1 & 2 & 1 & 2 & 2 & 2 & 1 & 2 & 2 & 2 & 2 & 2 & 2 & 2 \\ \hline
		\rowcolor{Gray}
			$\mathbf{h}_{28}$ & 0.0720 & 0.0488 & 0.0803 & 0.0488 & 0.0803 & 0.0803 & 0.0880 & 0.0487 & 0.0803 & 0.0822 & 0.0875 & 0.0822 & 0.0875 & 0.0833 & 0.0886  \\ \hline			$\rho_{29}$ & 1 & 2 & 2 & 1 & 2 & 2 & 2 & 1 & 2 & 2 & 2 & 2 & 2 & 2 & 2 \\ \hline
		\rowcolor{Gray}
			$\mathbf{h}_{29}$ & 0.0434 & 0.0644 & 0.0716 & 0.0434 & 0.0733 & 0.0716 & 0.0779 & 0.0435 & 0.0722 & 0.0716 & 0.0782 & 0.0723 & 0.0747 & 0.0782 & 0.0793  \\ \hline
		$\rho_{30}$ & 1 & 1 & 2 & 2 & 2 & 2 & 2 & 1 & 2 & 2 & 2 & 2 & 2 & 2 & 2 \\ \hline
		\rowcolor{Gray}
			$\mathbf{h}_{30}$ & 0.0468 & 0.0469 & 0.0781 & 0.0693 & 0.0773 & 0.0773 & 0.0845 & 0.0469 & 0.0791 & 0.0780 & 0.0806 & 0.0773 & 0.0842 & 0.0845 & 0.0856  \\ \hline
		$\rho_{31}$ & 1 & 1 & 2 & 1 & 2 & 2 & 2 & 2 & 2 & 2 & 2 & 2 & 2 & 2 & 2 \\ \hline
		\rowcolor{Gray}
			$\mathbf{h}_{31}$ & 0.0469 & 0.0469 & 0.0771 & 0.0468 & 0.0790 & 0.0790 & 0.0801 & 0.0693 & 0.0772 & 0.0772 & 0.0846 & 0.0772 & 0.0841 & 0.0841 & 0.0852  \\ \hline			$\rho_{32}$ & 2 & 1 & 3 & 1 & 3 & 2 & 3 & 1 & 3 & 2 & 3 & 2 & 3 & 3 & 3 \\ \hline
		\rowcolor{Gray}
			$\mathbf{h}_{32}$ & 0.2586 & 0.1313 & 0.3408 & 0.1313 & 0.3408 & 0.2490 & 0.3707 & 0.1313 & 0.3408 & 0.2490 & 0.3707 & 0.2490 & 0.3707 & 0.3512 & 0.3826  \\ \hline			$\rho_{33}$ & 1 & 2 & 3 & 1 & 2 & 3 & 3 & 1 & 2 & 3 & 3 & 2 & 3 & 3 & 3 \\ \hline
		\rowcolor{Gray}
			$\mathbf{h}_{33}$ & 0.1291 & 0.2542 & 0.3349 & 0.1291 & 0.2446 & 0.3349 & 0.3641 & 0.1291 & 0.2446 & 0.3349 & 0.3641 & 0.2446 & 0.3449 & 0.3641 & 0.3757  \\ \hline			$\rho_{34}$ & 1 & 1 & 2 & 2 & 3 & 3 & 3 & 1 & 2 & 2 & 3 & 3 & 3 & 3 & 3 \\ \hline
		\rowcolor{Gray}
			$\mathbf{h}_{34}$ & 0.1281 & 0.1280 & 0.2426 & 0.2522 & 0.3322 & 0.3322 & 0.3611 & 0.1281 & 0.2426 & 0.2426 & 0.3422 & 0.3322 & 0.3611 & 0.3611 & 0.3726  \\ \hline			$\rho_{35}$ & 1 & 1 & 2 & 1 & 2 & 2 & 3 & 2 & 3 & 3 & 3 & 3 & 3 & 3 & 3 \\ \hline
		\rowcolor{Gray}
			$\mathbf{h}_{35}$ & 0.1282 & 0.1282 & 0.2428 & 0.1282 & 0.2428 & 0.2428 & 0.3424 & 0.2524 & 0.3325 & 0.3325 & 0.3615 & 0.3325 & 0.3615 & 0.3615 & 0.3730  \\ \hline
	\end{tabular}
\end{table}

\subsection{Distribution with $\mathbb{I}(t(\mathbf{h}^{\mathrm{{ti}}}))=$-0.092499}\label{sec:app1}
{\tiny
\begin{align*}
p_{0000}=0.000012846134551, p_{0100}=0.000000000301036,   p_{0200}=0.000000004165347, p_{0300}=0.000001352300178,\\ 
p_{0400}= 0.000000878116619,  p_{1000}=0.000001269150960,  p_{1100}=0.000000000005073, p_{1200}=0.000000000013344,\\  p_{1300}=0.000000002478160,  p_{1400}=0.000000001625758,  p_{2000}=0.000000016909568,  p_{2100}=0.000000000000003,\\  p_{2200}=0.000000000005037,  p_{2300}=0.000000000182980,  p_{2400}=0.000000000118199,  p_{3000}=0.000387689258027,\\  p_{3100}=0.000000001904826,  p_{3200}=0.000000141961943,  p_{3300}=0.000000883109382,  p_{3400}=0.000000573732510,\\  p_{4000}=0.003112962301701,  p_{4100}=0.000000015294475,  p_{4200}=0.000001143842934,  p_{4300}=0.000007089870694,\\  p_{4400}=0.000004626359241,  p_{0010}=0.001619633838341,  p_{0110}=0.093591397567733,  p_{0210}=0.000005413323836,\\  p_{0310}=0.000000000000295,  p_{0410}=0.000000000000222,  p_{1010}=0.001621729575016,  p_{1110}=0.001610147592648,\\  p_{1210}=0.000000001784253,  p_{1310}=0.000000000000003,  p_{1410}=0.000000000000001,  p_{2010}=0.000000000000001,\\  p_{2110}=0.000000000000445,  p_{2210}=0.000000000000001,  p_{3010}=0.000000000206447,  p_{3110}=0.000000613214019,\\  p_{3210}=0.000000000000004,  p_{4010}=0.000000001656310, p_{4110}=0.000004919105728,   p_{4210}=0.000000000000055,\\ p_{0020}=0.000000002117842,  p_{0120}=0.000090927231165,  p_{0220}=0.000015744231593,  p_{0320}=0.000000020603595,\\ p_{0420}=0.000000013219636,  p_{1020}=0.000000121026957, p_{1120}=0.024608797507497,  p_{1220}=0.000091981467264,\\ p_{1320}=0.000005677207854,  p_{1420}=0.000003691676566, p_{2020}=0.000000000000013,  p_{2120}=0.000000120577505,\\ p_{2220}=0.000000002132127,  p_{2320}=0.000000000024431, p_{2420}=0.000000000015856,   p_{3020}=0.000000000004537,\\ p_{3120}=0.000001029356910,  p_{3220}=0.000000003683362, p_{3320}=0.000000000007856,  p_{3420}=0.000000000005131,\\ p_{4020}=0.000000000035430,  p_{4120}=0.000008269581152, p_{4220}=0.000000029815681,  p_{4320}=0.000000000062828,\\ p_{4420}=0.000000000041303,  p_{0030}=0.000079361827355, p_{0130}=0.000000001849276,  p_{0230}=0.000000025983915,\\  p_{0330}=0.000008345672352,  p_{0430}=0.000005432402419, p_{1030}=0.000007817664095,  p_{1130}=0.000000000031659,\\ p_{1230}=0.000000000086481,  p_{1330}=0.000000015330644, p_{1430}=0.000000010012229,  p_{2030}=0.000000104510163,\\ p_{2130}=0.000000000000006,  p_{2230}=0.000000000030881, p_{2330}=0.000000001128239,  p_{2430}=0.000000000731944,\\ p_{3030}=0.002397876230926,  p_{3130}=0.000000011797010, p_{3230}=0.000000881508050,  p_{3330}=0.000005488281524,\\
p_{3430}=0.000003565927666,  p_{4030}=0.019190048010682, p_{4130}=0.000000094539083,  p_{4230}=0.000007067061506,\\ p_{4330}=0.000043931012281,  p_{4430}=0.000028572774306, p_{0040}=0.094340650560348,  p_{0140}=0.001631865128852,\\ p_{0240}=0.000000000000403,  p_{0340}=0.000003436005505,  p_{0440}=0.000002239820032, p_{1040}=0.000005331058685,\\ p_{1140}=0.000000001775116,  p_{1340}=0.000000000000039,  p_{1440}=0.000000000000026, p_{2040}=0.000000000000498,\\  p_{2140}=0.000000000000002, p_{3040}=0.000183634926005,  p_{3140}=0.000181812651107,   p_{3340}=0.000000000131451,\\  p_{3440}=0.000000000085363, p_{4040}=0.001469168541679,  p_{4140}=0.001455298777297, p_{4240}=0.000000000000003,\\  p_{4340}=0.000000001049293, p_{4440}=0.000000000685053,  p_{0001}=0.000000000014370, p_{0101}=0.000000000005101,\\  p_{0201}=0.000001272968642, p_{0301}=0.000000002556229,  p_{0401}=0.000000001675382,  p_{1001}=0.000000004339229,\\  p_{1101}=0.000000000298701, p_{1201}=0.000012856874276,  p_{1301}=0.000001354910658, p_{1401}=0.000000877484807,\\  p_{2001}=0.000001289017977, p_{2101}=0.000000017230495,  p_{2201}=0.003492332326638, p_{2301}=0.000007986034205,\\  p_{2401}=0.000005195934596, p_{3001}=0.000000000000623,  p_{3201}=0.000000001845713,  p_{3301}=0.000000000020622,\\  p_{3401}=0.000000000013201, p_{4001}=0.000000000004456,  p_{4101}=0.000000000000001, p_{4201}=0.000000015038067,\\  p_{4301}=0.000000000163836, p_{4401}=0.000000000104890,  p_{0011}=0.000000001787423, p_{0111}=0.001611505670007,\\  p_{0211}=0.001625151140990, p_{0311}=0.000000000000001,  p_{0411}=0.000000000000002, p_{1011}=0.000005430254468,\\   p_{1111}=0.093637527230351, p_{1211}=0.001615642248607,  p_{1311}=0.000000000000319, p_{1411}=0.000000000000195,\\  p_{2011}=0.000000000000060, p_{2111}=0.000005551162205,  p_{2211}=0.000000001850074, p_{2411}=0.000000000000001,\\   p_{3111}=0.000000000000053, p_{4111}=0.000000000000408,  p_{4211}=0.000000000000005, p_{0121}=0.000000001767046,\\ p_{0221}=0.000005323994495,  p_{0321}=0.000000000000032, p_{0421}=0.000000000000038,  p_{1021}=0.000000000000480,\\ p_{1121}=0.001626361371904,   p_{1221}=0.094161151405879, p_{1321}=0.000003443402741,  p_{1421}=0.000002226347281,\\ p_{2021}=0.000000000000002,  p_{2121}=0.001634965798292, p_{2221}=0.001650144379024,  p_{2321}=0.000000001187669,\\ p_{2421}=0.000000000774893,  p_{3121}=0.000000000000002,  p_{3221}=0.000000000000056, p_{3421}=0.000000000000001,\\    p_{4121}=0.000000000000003, p_{4221}=0.000000000000475,  p_{4321}=0.000000000000001,  p_{0031}=0.000000000087713,\\ p_{0131}=0.000000000031965,  p_{0231}=0.000007859341096, p_{0331}=0.000000015832013,  p_{0431}=0.000000010238436,\\ p_{1031}=0.000000026526025,  p_{1131}=0.000000001845107, p_{1231}=0.000079121315460,  p_{1331}= 0.000008322716936,\\ p_{1431}=0.000005417381121,  p_{2031}=0.000007969071794, p_{2131}=0.000000106344016,  p_{2231}=0.021558780509219,\\ p_{2331}=0.000049322992186,  p_{2431}=0.000032050062484, p_{3031}=0.000000000003513,  p_{3131}=0.000000000000006,\\ p_{3231}=0.000000011559789,  p_{3331}=0.000000000123992, p_{3431}=0.000000000080506,  p_{4031}=0.000000000028006,\\  p_{4131}=0.000000000000011,  p_{4231}=0.000000092408526, p_{4331}=0.000000000996030,  p_{4431}=0.000000000648610,\\ p_{0041}=0.000092358786737,  p_{0141}=0.024712495134647, p_{0241}=0.000000121437730,  p_{0341}=0.000005680667166,\\ p_{0441}=0.000003688700569,  p_{1041}=0.000015755360355, p_{1141}=0.000091277644612,  p_{1241}=0.000000002125769,\\ p_{1341}=0.000000019826459,   p_{1441}=0.000000012946325, p_{2041}=0.000000032341419,  p_{2141}=0.000009292856848,\\ p_{2241}=0.000000000038731,  p_{2341}=0.000000000069148, p_{2441}=0.000000000044807,  p_{3041}=0.000000000236053,\\ p_{3141}=0.000000013361490,  p_{3341}=0.000000000002687,  p_{3441}=0.000000000001671, p_{4041}=0.000000001890320,\\  p_{4141}=0.000000106802609,  p_{4241}=0.000000000000005,  p_{4341}=0.000000000020935, p_{4441}=0.000000000013507,\\  p_{0002}=0.000000000267215,  p_{0202}=0.000000000000011, p_{0302}=0.000000459990980,  p_{0402}=0.000000300424887,\\ p_{1002}=0.000000000000001,  p_{1302}=0.000000000000057, p_{1402}=0.000000000000021,  p_{2002}=0.000232627467886,\\ p_{2102}=0.000000000000001,  p_{2202}=0.000000000277738, p_{2302}=0.000145056855474,  p_{2402}=0.000094139813988,\\ p_{3002}=0.000026310786346,  p_{3102}=0.000000000000010, p_{3202}=0.000000086839960,  p_{3302}=0.000898922432433,\\ p_{3402}=0.000585721863654,  p_{4002}=0.000211062688232, p_{4102}=0.000000000000048,  p_{4202}=0.000000692942053,\\ p_{4302}=0.007222885632422,  p_{4402}=0.004701152379961, p_{0012}=0.024829648337754,  p_{0112}=0.000091458083901,\\ p_{0212}=0.000009190715664,  p_{0312}=0.000000073900687, p_{0412}=0.000000047562884,  p_{1012}=0.000000123049530,\\ p_{1112}=0.000000002149085,  p_{1212}=0.000000000039218, p_{1312}=0.000000000000009,  p_{1412}=0.000000000000008,\\ p_{2012}=0.000009375814717,  p_{2112}=0.000000032765477, p_{2212}=0.000000000110840,  p_{2312}=0.000000000023901,\\ p_{2412}=0.000000000015363,  p_{3012}=0.000010372097811, p_{3112}=0.000001761200481,  p_{3212}=0.000000003551441,\\ p_{3312}=0.000000000145579,  p_{3412}=0.000000000093563, p_{4012}=0.000083252168163,  p_{4112}=0.000014126709971,\\ 
p_{4212}=0.000000028211367,  p_{4312}=0.000000001166270, p_{4412}=0.000000000753934,  p_{0022}=0.000000000034831,\\ p_{0122}=0.000000000093916,  p_{0222}=0.000000028474896, p_{0322}=0.000005517472236,  p_{0422}=0.000003600337473,\\ p_{1022}=0.000000000000006,  p_{1122}=0.000000000035083, p_{1222}=0.000000002156298,  p_{1322}=0.000000074757363,
\end{align*}
\begin{align*}
p_{1422}=0.000000048756891,  p_{2022}=0.000000122061025, p_{2122}=0.000009093930833,  p_{2222}=0.000093228954856,\\ p_{2322}=0.015304222047613,  p_{2422}=0.009967681536496, p_{3022}=0.000000000237613,  p_{3122}=0.000000003167821,\\ p_{3222}=0.000001772039889,  p_{3322}=0.000006329774200, p_{3422}=0.000004126086208,   p_{4022}=0.000000001915268,\\ p_{4122}=0.000000025360348,  p_{4222}=0.000014254572572, p_{4322}=0.000050918007063,  p_{4422}=0.000033152056951,\\ p_{0032}=0.000000001650025,  p_{0232}=0.000000000000057,  p_{0332}=0.000002854995278, p_{0432}=0.000001863772864,\\
p_{1032}=0.000000000000003,  p_{1232}=0.000000000000001, p_{1332}=0.000000000000305,  p_{1432}=0.000000000000143,\\ p_{2032}=0.001434696410495,  p_{2132}=0.000000000000001, p_{2232}=0.000000001710196,  p_{2332}=0.000894951036323,\\ p_{2432}=0.000583860656217,  p_{3032}=0.000161727409343, p_{3132}=0.000000000000046,  p_{3232}=0.000000538140698,\\ p_{3332}=0.005557834401157,  p_{3432}=0.003622047218672, p_{4032}=0.001302817560730,  p_{4132}=0.000000000000314,\\ p_{4232}=0.000004298032295,  p_{4332}=0.044537461873833, p_{4432}=0.029026666003322, p_{0042}=0.001670334744037,\\ p_{0142}=0.000000001830798,  p_{0242}=0.000000000000003, p_{0342}=0.001002413475281,  p_{0442}=0.000653486467145,\\ p_{1042}=0.000000000000533,  p_{1142}=0.000000000000001, p_{2042}=0.000005713039484, p_{2142}=0.000000000000067,\\  p_{2342}=0.000000001242140,  p_{2442}=0.000000000797143, p_{3042}=0.010588925045547,  p_{3142}=0.000000601458506,\\ p_{3242}=0.000000000000046,  p_{3342}=0.000114342942426, p_{3442}=0.000074603916825,  p_{4042}=0.084728572731005,\\ p_{4142}=0.000004818701252,  p_{4242}=0.000000000000500, p_{4342}=0.000912767278616,  p_{4442}=0.000594931241561,\\ 
p_{0103}=0.000000000000001,  p_{0303}=0.000000000000023, p_{0403}=0.000000000000009,  p_{1003}=0.000000000000005,\\ p_{1203}=0.000000000171523, p_{1303}=0.000000301697471,  p_{1403}=0.000000197324378, p_{2003}=0.000000510614681,\\  p_{2103}=0.000000000000046, p_{2203}=0.000154907487031,  p_{2303}=0.005280821394789, p_{2403}=0.003434706714543,\\  p_{3003}=0.000000000019767, p_{3203}=0.000016711837566,  p_{3303}=0.000010431816580, p_{3403}=0.000006752737601,\\  p_{4003}=0.000000000160701, p_{4203}=0.000134555622746, p_{4303}=0.000084143939140,  p_{4403}=0.000054593229675,\\ p_{0013}=0.000000000024802,  p_{0113}=0.000000001390365, p_{0213}=0.000000079693490,  p_{0313}=0.000000000000007,\\ p_{0413}=0.000000000000005,  p_{1013}=0.000005984399516, p_{1113}=0.000059606899104,  p_{1213}=0.016173546540424,\\ p_{1313}=0.000000047734427,  p_{1413}=0.000000030981318, p_{2013}=0.000000020416962,  p_{2113}=0.000010365792628,\\ p_{2213}=0.000061236848700,  p_{2313}=0.000000000862016, p_{2413}=0.000000000553032,  p_{3013}=0.000000000007822,\\ p_{3113}=0.000000002324035,  p_{3213}=0.000000671721543, p_{3313}=0.000000000001641,  p_{3413}=0.000000000001049,\\ p_{4013}=0.000000000063689,  p_{4113}=0.000000018449481, p_{4213}=0.000005419900711, p_{4313}= 0.000000000013052,\\ p_{4413}=0.000000000008805, p_{0223}=0.000000000000306, p_{0323}=0.000000000000001,  p_{0423}=0.000000000000001,\\
p_{1023}=0.000000000000001,  p_{1123}=0.000000001175285, p_{1223}=0.001080741761477,  p_{1323}=0.000652502763271,\\ p_{1423}=0.000424290949571,  p_{2023}=0.000000000000314, p_{2123}=0.000003486269401,  p_{2223}=0.062021417579880,\\ p_{2323}=0.000668760548162,  p_{2423}=0.000435160996265, p_{3123}=0.000000000000006, p_{3223}=0.000000410659423,\\ p_{3323}=0.000000000087749, p_{3423}=0.000000000057440,  p_{4123}=0.000000000000044,  p_{4223}=0.000003307432883,\\ p_{4323}=0.000000000712676,  p_{4423}=0.000000000462515, p_{0233}=0.000000000000001,  p_{0333}=0.000000000000159,\\ p_{0433}=0.000000000000099, p_{1033}=0.000000000000050,  p_{1233}=0.000000001064245, p_{1333}=0.000001866426432,\\  p_{1433}=0.000001211269238, p_{2033}=0.000003143167796,  p_{2133}=0.000000000000235, p_{2233}=0.000954958581357,\\  p_{2333}=0.032604207478090, p_{2433}=0.021209434584384,  p_{3033}=0.000000000124358, p_{3133}=0.000000000000001,\\  p_{3233}=0.000103394689450, p_{3333}=0.000064452929897,  p_{3433}=0.000041918631639, p_{4033}=0.000000000987985,\\  p_{4133}=0.000000000000001, p_{4233}=0.000830280563733,  p_{4333}=0.000518665216863, p_{4433}=0.000336753185843,\\  p_{0043}=0.000000001400638, p_{0143}=0.000000000023344,  p_{0243}=0.000000000000006, p_{0343}=0.000000048553203,\\  p_{0443}=0.000000031625775, p_{1043}=0.000000018987374,  p_{1143}=0.000000000062721, p_{1243}=0.000000000023304,\\  p_{1343}=0.000003610386387, p_{1443}=0.000002346141896,  p_{2043}=0.000010427453537, p_{2143}=0.000000019023682,\\  p_{2243}=0.000000001404738, p_{2343}=0.000037184160438,  p_{2443}=0.000024220031036, p_{3043}=0.000006731473764,\\  p_{3143}=0.000000659039552, p_{3243}=0.000000008831768,  p_{3343}=0.001102868429136, p_{3443}=0.000717811356179,\\  p_{4043}=0.000054007910233, p_{4143}=0.000005286400245,  p_{4243}=0.000000070751009, p_{4343}=0.008846180025935,\\  p_{4443}=0.005760459413997, p_{0204}=0.000000000000001,  p_{0304}=0.000000000000032, p_{0404}=0.000000000000009,\\ 
p_{1004}=0.000000000000003, p_{1204}=0.000000000092017, p_{1304}=0.000000162088415,  p_{1404}=0.000000107120011,\\ p_{2004}=0.000000273881081,  p_{2104}=0.000000000000020, p_{2204}=0.000083380004818,  p_{2304}=0.002836651578416,\\ p_{2404}=0.001847818727558,  p_{3004}=0.000000000010860, p_{3204}=0.000009018108453, p_{3304}=0.000005600939106,\\  p_{3404}=0.000003644783274, p_{4004}=0.000000000085077,  p_{4104}=0.000000000000001, p_{4204}=0.000072325656139,\\  p_{4304}=0.000045022135794, p_{4404}=0.000029380448792,  p_{0014}=0.000000000013613, p_{0114}=0.000000000747923,\\  p_{0214}=0.000000043189142, p_{0314}=0.000000000000003,  p_{0414}=0.000000000000001, p_{1014}=0.000003217573451,\\  p_{1114}=0.000031958500711, p_{1214}=0.008696190783716,  p_{1314}=0.000000025915307, p_{1414}=0.000000016662202,\\  p_{2014}=0.000000010969771, p_{2114}=0.000005561372417,  p_{2214}=0.000032848973935, p_{2314}=0.000000000456750,\\  p_{2414}=0.000000000296645, p_{3014}=0.000000000004263,  p_{3114}=0.000000001220998, p_{3214}=0.000000361006163,\\  p_{3314}=0.000000000000903, p_{3414}=0.000000000000598,  p_{4014}=0.000000000033723, p_{4114}=0.000000009906092,\\  p_{4214}=0.000002891784422, p_{4314}=0.000000000007273,  p_{4414}=0.000000000004731, p_{0224}=0.000000000000182,\\  p_{0324}=0.000000000000001, p_{0424}=0.000000000000003,  p_{1124}=0.000000000631430, p_{1224}=0.000581281437928,\\ p_{1324}=0.000350268288064,  p_{1424}=0.000228482006076, p_{2024}=0.000000000000186, p_{2124}=0.000001867710330,\\ p_{2224}=0.033351170539991,  p_{2324}=0.000358916594735, p_{2424}=0.000233432479259,  p_{3124}=0.000000000000002,\\  p_{3224}=0.000000221483852, p_{3324}=0.000000000047686,  p_{3424}=0.000000000030493,  p_{4124}=0.000000000000016,\\ p_{4224}=0.000001772191406,  p_{4324}=0.000000000377785, p_{4424}=0.000000000249065, p_{0334}=0.000000000000073,\\  p_{0434}=0.000000000000068, p_{1034}=0.000000000000020, p_{1234}=0.000000000570750,  p_{1334}=0.000000998333661,\\ p_{1434}=0.000000652998931,  p_{2034}=0.000001691265398, p_{2134}=0.000000000000135,  p_{2234}=0.000513886049713,\\ p_{2334}=0.017497860689953,  p_{2434}=0.011423419875122, p_{3034}=0.000000000065730,  p_{3234}=0.000055656446240,\\  p_{3334}=0.000034592870576, p_{3434}=0.000022600420150,  p_{4034}=0.000000000533660, p_{4134}=0.000000000000002,\\  p_{4234}=0.000446007554828, p_{4334}=0.000278534966469,  p_{4434}=0.000181276825181, p_{0044}=0.000000000757445,\\  p_{0144}=0.000000000012723, p_{0244}=0.000000000000006,  p_{0344}=0.000000026206720, p_{0444}=0.000000017119455,\\  p_{1044}=0.000000010058038, p_{1144}=0.000000000033889,  p_{1244}=0.000000000012354, p_{1344}=0.000001934288033,\\ p_{1444}=0.000001265899157, p_{2044}=0.000005620764724,  p_{2144}=0.000000010306073,  p_{2244}=0.000000000760747,\\  p_{2344}=0.000020042413662, p_{2444}=0.000013046091217,  p_{3044}=0.000003626184776,  p_{3144}=0.000000354330068,\\  p_{3244}=0.000000004771155, p_{3344}=0.000593203069042,  p_{3444}=0.000387349247884,  p_{4044}=0.000029042156624,\\  p_{4144}=0.000002832754100, p_{4244}=0.000000038340538,  p_{4344}=0.004746348339701, p_{4444}=0.003098320475986,
\end{align*}
and else zero.
}

\end{document}